\documentclass[journal]{IEEEtran}  % Comment this line out
\IEEEoverridecommandlockouts                              % This command is only
\usepackage{graphicx}
\usepackage{mathtools, cuted}
\usepackage{booktabs}
\usepackage{comment}
\usepackage{cite}

\usepackage{graphics}
\usepackage[table,xcdraw]{xcolor}
\usepackage{multicol}
\usepackage{newunicodechar}
\newunicodechar{⌈}{\lceil}
\usepackage{multirow}
\usepackage[linesnumbered,ruled,vlined]{algorithm2e}
\SetKwRepeat{Do}{do}{while}%
\usepackage{amsmath,amssymb}
\usepackage{caption}
\usepackage{paralist}
\usepackage{dblfloatfix}
\usepackage{subcaption}
\usepackage[utf8]{inputenc}
\usepackage{textcomp}
\usepackage{newunicodechar}
\newunicodechar{−}{\textminus}
\usepackage{graphicx}
\usepackage{setspace}
\usepackage{wrapfig}
\usepackage[hyphens]{url}
\usepackage{amsthm}
\usepackage{relsize}
\usepackage{amssymb}% http://ctan.org/pkg/amssymb
\usepackage{pifont}% http://ctan.org/pkg/pifont
\usepackage{float}
\usepackage{graphicx}
\usepackage{xcolor}
\usepackage[table]{xcolor}
\usepackage{caption}
\usepackage{booktabs}
\usepackage{array}
\usepackage{eso-pic}
\usepackage{xcolor}
\usepackage{soul}
\usepackage{url}

\usepackage[most]{tcolorbox}
\usepackage{hyperref}
\usepackage[table,xcdraw]{xcolor}

\newcommand{\name}{\texttt{DISARM}}
\newtcolorbox{takeaway}[1][]{
  enhanced, breakable,
  colback=black!4, colframe=black!60,
  boxrule=0.5pt, arc=0.6mm,
  left=6pt,right=6pt,top=6pt,bottom=6pt,
  title={}, fonttitle=\bfseries,
  #1
}
\makeatletter
\newcommand{\multiline}[1]{%
  \begin{tabularx}{\dimexpr\linewidth-\ALG@thistlm}[t]{@{}X@{}}
    #1
  \end{tabularx}
}
\makeatother

\title{
{\normalfont\footnotesize
\textcopyright~2026 IEEE. Personal use of this material is permitted.
Permission from IEEE must be obtained for all other uses, in any current or
future media, including reprinting/republishing this material for advertising
or promotional purposes, creating new collective works, for resale or
redistribution to servers or lists, or reuse of any copyrighted component of
this work in other works.

The definitive version of this article has been accepted for publication in the
IEEE Transactions on Information Forensics and Security (TIFS).
}
DISARM: Target Electronic Device Informed Mitigation of Software Runtime Side-Channel Vulnerabilities
}

\author{
\centering
Tasneem Suha$^{1}$, Tanzim Mahfuz$^{1}$, Rima Asmar Awad$^{2}$, Prabuddha Chakraborty$^{1,*}$\\[0.8em]
$^{1}$Department of Electrical \& Computer Engineering, University of Maine, Orono, ME, USA\\
$^{2}$Oak Ridge National Laboratory, Oak Ridge, TN, USA\\[0.4em]
\texttt{\{tasneem.suha, tanzim.mahfuz, prabuddha\}@maine.edu}\\
\texttt{rima.awad@ornl.gov}\\[0.4em]
$^{*}$\textit{Corresponding Author}
}
\date{}

\begin{document}

\maketitle

\thispagestyle{empty}
\pagestyle{empty}

%%%%%%%%%%%%%%%%%%%%%%%%%%%%%%%%%%%%%%%%%%%%%%%%%%%%%%%%%%%%%%%%%%%%%%%%%%%%%%%%
% Such an attack can lead to the leakage of encryption keys, propriety additive manufacturing print pathways, and sensitive data in-compute. 
% However, such software fixing typically do not consider the specific hardware the software is supposed to execute on in the field, leading to over/under fixing of the vulnerability. 
% Recent works such as HASTE, highlights this issue in timing Side-channel mitigation frameworks and proposed a more tailored methodology to quantify timing Side-channel vulnerabilities while considering the hardware architecture of the target system using a simulation frameworks (RISC-V BOOM). However, such simulations are often inaccurate particularly when dealing with timing values. 
\begin{abstract}
Program runtime/timing attacks exploit variations in a program's execution times to extract sensitive information from the program (e.g. encryption keys, sensitive variable data, intellectual property). State-of-the-art solutions to runtime side-channel attacks attempt to balance the execution time of the sensitive code for different control flow paths to eliminate the timing leakage. However, during the mitigation process, most techniques do not consider the underlying hardware/device on which the target program is supposed to run on. This can lead to over-fixing (unnecessary extra operations), under-fixing (not solving the imbalance properly), and even failures. We propose \name, a joint hardware-software methodology (unlike any existing solution) for mitigating runtime side-channel vulnerabilities that utilizes timing values from real embedded devices to generate targeted software fixes. We implement \name~to support C/C++/Java source codes and validate it across 22 standard benchmarks. \name~outperforms state-of-the-art solutions such as PENDULUM and DifFuzzAR in terms of execution time overhead, code size overhead, and correctness on five different embedded/edge devices.   

% on four embedded devices using six software benchmarks and observe significant post-fixed codes' performance advantage compared to constant-time programming.  

% follow a constant-time programming approach which can lead to fix failures (due to program complexity) and over-fixing/under

% without much consideration given to the underlying hardware). 

% HASTE-FIX can automatically patch a given software for a given electronic hardware (on which the software will run) while also being mindful of the capabilities of a given attacker model. This allows \name~ to mitigate the vulnerabilities appropriately without over/under fixing leading to increased edge energy efficiency and performance.                   

% , potentially leading to the disclosure of confidential information and system vulnerabilities. This inadequacy in traditional mitigation strategies arises because timing variations are inherently influenced by the underlying hardware architecture. In this paper, we propose a highly parameterized timing side-channel mitigation framework which will consider the attacker timing sensitivity and utilize the embedded hardware to identify and mitigate timing side-channel vulnerability in a more targeted manner. We verify our proposed framework on six different sensitive software subroutines that are widely used in many mission-critical systems (e.g. additive manufacturing, health care, cryptosystem etc). The verification was conducted across five different edge devices to demonstrate the framework’s adaptability and effectiveness in diverse systems. 

\end{abstract}


%%%%%%%%%%%%%%%%%%%%%%%%%%%%%%%%%%%%%%%%%%%%%%%%%%%%%%%%%%%%%%%%%%%%%%%%%%%%%%%%

% “They’re not that hard to mitigate” Survey
% "You’re Too Slow!"

\AddToShipoutPictureFG*{%
  \AtPageLowerLeft{%
    \raisebox{0.1in}{%
      \makebox[\paperwidth]{%
        \colorbox{white}{%
          \parbox[b]{0.890\paperwidth}{%
            \footnotesize
            \centering
            \textit{Notice: This manuscript has been authored by UT-Battelle, LLC,
            under contract DE-AC05-00OR22725 with the US Department of Energy (DOE).
            The US government retains and the publisher, by accepting the article for
            publication, acknowledges that the US government retains a nonexclusive,
            paid-up, irrevocable, worldwide license to publish or reproduce the
            published form of this manuscript, or allow others to do so, for US
            government purposes. DOE will provide public access to these results of
            federally sponsored research in accordance with the DOE Public Access Plan
            \,(\url{https://www.energy.gov/doe-public-access-plan}).%
          }}%
        }%
      }%
    }%
  }%
}

\section{Introduction}
Low power embedded/edge devices are being widely used in different critical applications such as Industry 4.0, Healthcare, Surveillance, and Aerospace. These devices often handle sensitive data (e.g. additive manufacturing print pathways, encryption keys, patient information, sensitive biometric identifiers) making them a target for a wide range of adversaries \cite{hassan2019current,   devi2021side, zankl2021side, alladi2020consumer, atlam2020iot}. 
These adversaries can employ different types of cyber-attacks to compromise the device. Such attacks include tampering, power side-channel attacks, firmware reverse engineering, and timing side-channel attacks \cite{standaert2010introduction,spreitzer2017systematic,zhou2005side,kopf2007information,lyu2018survey}. In this work, we focus on timing side-channel attack specifically based on program runtime that is generally more challenging to address due to heavy reliance on both the software and the hardware components of the embedded/edge devices.

\begin{figure}[!ht]
\centering

\includegraphics[width=\columnwidth]{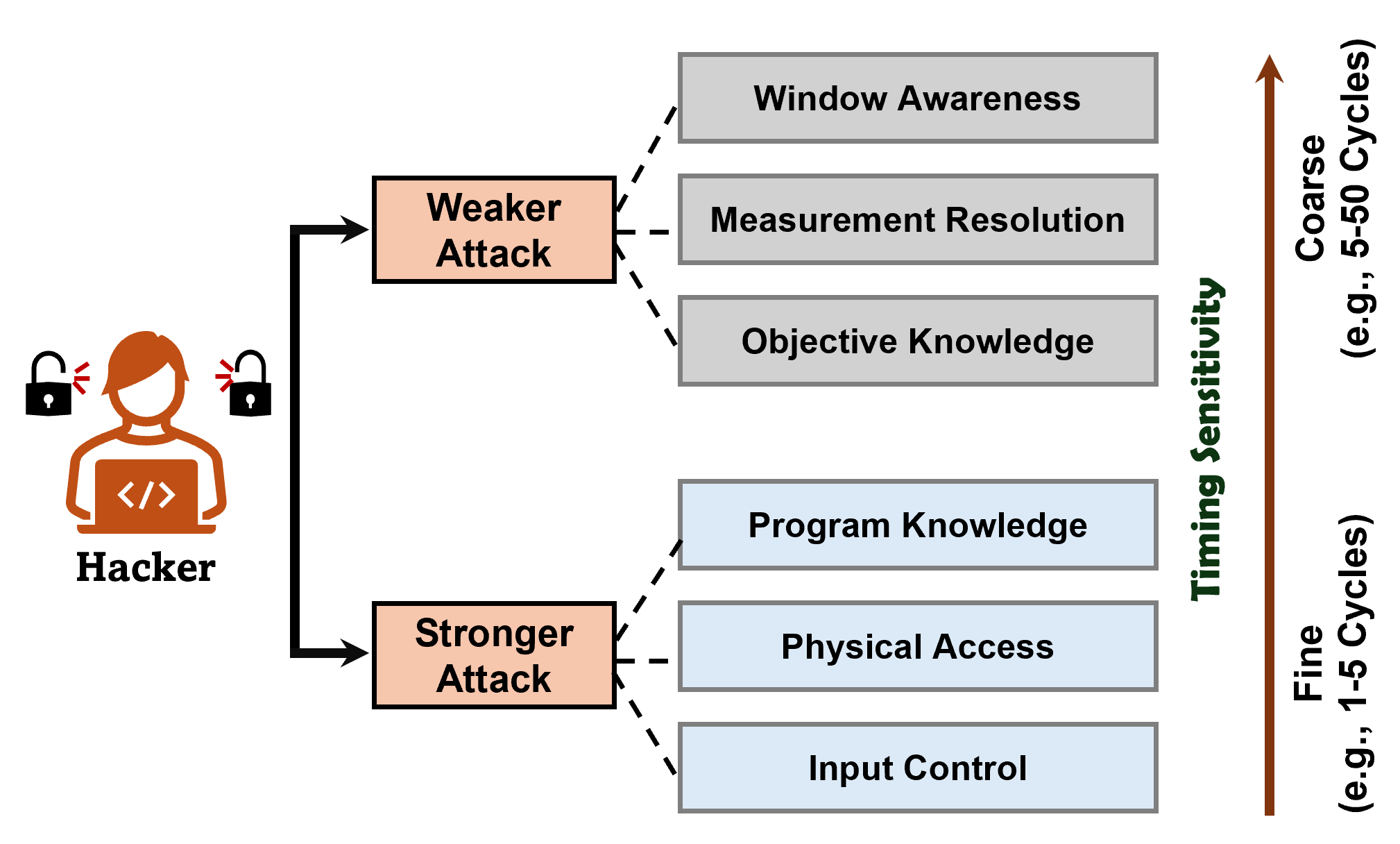}
\captionsetup{justification=centering}
\caption{Threat Model: Strength of the attacker depends on the level of access/knowledge.\label{fig:threat}}

% \vspace{-0.25in}
\end{figure}
%\centering
\begin{figure*}[]
\centering

\includegraphics[width=\linewidth]{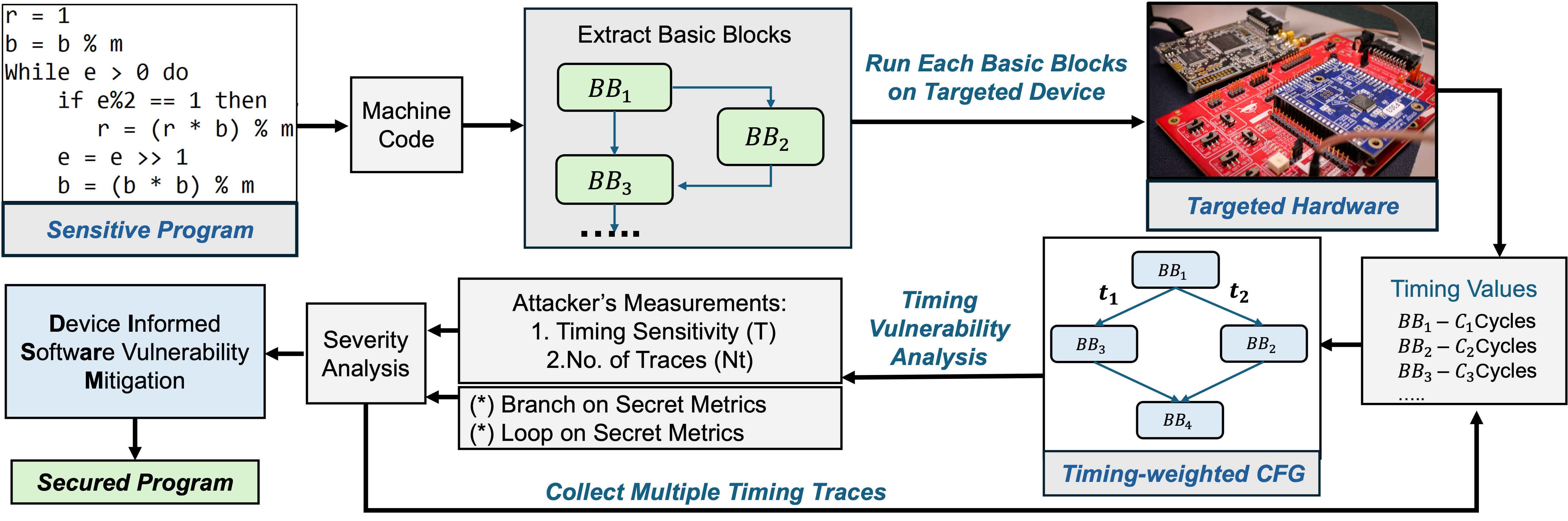}
\caption{\small{Overview of \name~: Hardware-in-the-loop and specific threat model tailored mitigation of software timing vulnerabilities.}\label{fig:haste-fix}}

\vspace{-0.2in}
\end{figure*}

A timing/runtime side-channel attack involves the attacker measuring and analyzing the execution time of a given program to determine the values of sensitive data being processed by the program \cite{zhang2024timing,neve2006refined,chevallier2004low,rohatgi2009improved}. This execution time is dependent not only on the software but also on the underlying electronic hardware due to differences in pipelining, hazard control, speculation, and wider architectural differences \cite{Hardware}. The level of threat also depends on the effectiveness of the attacker in terms of measuring the execution time of the code (see Fig.~\ref{fig:threat}). For example, an attacker having physical access to the device (a possibility because these devices are often deployed in open areas) may be able to obtain almost cycle-accurate estimation of the execution time \cite{li2021chronos, crosby2009opportunities, brumley2011remote, danev2010attacks,weiss2012cache}. In another threat model, the attacker may have only limited over-the-network access to the device and hence the execution time measurement may be less accurate\cite{brumley2005remote}. Hence, a timing side-channel mitigation framework must consider software, hardware, and also precise threat model to generate efficient and effective solutions. 

However, most state-of-the-art (e.g., DifFuzzAR \cite{diffuzzar}, PENDULUM \cite{pendulum}) runtime side-channel mitigation frameworks do not consider the underlying hardware and the specific attacker threat model during the mitigation/fixing process leading to gross over/under fixing \cite{watt2019ct,he2020ct, van2020timeless}. Recent works such as HASTE \cite{chakraborty2021haste} has looked into the quantification of different timing side-channel vulnerabilities based on a given software, hardware (only simulation), and the attacker threat model. However, HASTE was designed to only detect (no mitigation) timing side-channel vulnerabilities and utilized the Berkeley Out-of-Order Machine (BOOM)\cite{BoomCore} simulator for mimicking different hardware architectures (no real devices).

In this work, we propose \textbf{\name:} \textbf{\underline{D}}evice \textbf{\underline{I}}nformed \textbf{\underline{S}}oftw\textbf{\underline{a}}re Vulne\textbf{\underline{r}}ability \textbf{\underline{M}}itigation, a framework that addresses the main shortcomings of the state-of-the-art frameworks (e.g. HASTE, PENDULUM, DifFuzzAR) by: (1) Quantifying the timing side-channel vulnerabilities using real electronic devices/hardware in-the-loop; (2) Fixing/mitigating the discovered vulnerabilities based on the specific threat model and hardware, prioritizing both safety and increased system efficiency. We implement \name~and integrate it into the commercial embedded systems development flow for automatic detection and fixing of timing side-channel vulnerabilities. We evaluate \name~using 22 commonly used sensitive software functions on five Internet-of-Things (IoT) and edge/embedded devices: (1) Jetson Nano ARM Cortex-A57 MP Core processor; (2) Jetson Orin Nano ARM Cortex-A78AE processor; (3) Jetson AGX Xavier Carmel ARM v8.2; (4) Jetson Xavier NX Carmel ARM v8.2; (5) Raspberry Pi Quad-core Cortex-A72. Empirical results demonstrate the ability of \name~to precisely fix timing side-channel vulnerabilities while having minimal impact on system performance and energy efficiency (compared to constant-time programming solutions, DifFuzzAR \cite{diffuzzar}, and PENDULUM \cite{pendulum}). 

In summary, we make the following contributions:
\begin{enumerate}
    \item Developed an automated framework for performing hardware-in-the-loop evaluation of software timing side-channel vulnerabilities.
    \item Designed algorithms to optimally mitigate the discovered timing side-channel vulnerabilities for the specific threat model and the underlying hardware, limiting compute/energy overheads.
    \item Implemented the algorithms and framework as a highly parameterized automated tool that has been integrated into the C/C++/Java software development tool-chain.
    \item Evaluated the framework (in comparison to constant-time programming solutions, DifFuzzAR \cite{diffuzzar}, and PENDULUM \cite{pendulum}) using 22 software benchmarks (C/C++/Java) on 5 embedded devices.
    % \item Enhanced the framework's adaptability across multiple programming language environment, both C/C++ and Java.
\end{enumerate}

\section{Background}
Next, we will briefly discuss the considered threat models, key timing vulnerabilities, and related works in this area.

\subsection{Threat Model \& Underlying Hypotheses}
\label{threat model}
Based on the attacker's accessibility, two types of attacks can occur. To elevate the `weak attack', the attacker should have some access to the system. (1) The attacker understands the appropriate moments to commence and conclude the timing of the program. (2) The attacker can measure the time it takes to execute specific segments of the program. (3) The attacker possesses a fundamental objective of the system's purpose or goal. Based on this threat model, the attacker's timing sensitivity for weaker attacks will be:
% \vspace{-0.1in}
\begin{equation}
    \mathcal{T}_{base} = {w}_1(\mathcal{W}_a)+{w}_2(\mathcal{M}_r)+{w}_3(\mathcal{K}_{obj})
\end{equation}
where, $\mathcal{W}_a$ is the window awareness, $\mathcal{M}_r$ is the measurement resolution, $\mathcal{K}_{obj}$ is the objective knowledge. ${w}_1$, ${w}_2$, ${w}_3$ $\in$ $\mathbb{R}$ $\geq$ 0 are weights signifying the importance of each factor.
The attack can be made robust by the attacker if the attacker: (1) Possesses the software code or binary; (2) Holds physical access to the system; (3) Can manipulate specific inputs to the system.
For strong timing attacks: 
% \vspace{-0.8em}
\begin{equation}
    \mathcal{T}_{strong} = \frac{\mathcal{T}_{base}}{1+\alpha(\mathcal{K}_{prog}+\mathcal{P}+\mathcal{I})}
\end{equation}
% \vspace{-0.2in}
where $\mathcal{K}_{prog}$ is the full program knowledge, $\mathcal{P}$ is the complete physical access, and $\mathcal{I}$ is the partial but nontrivial input manipulation. $\alpha >$ 0, is a scaling factor controlling how drastically these strong-attacker attributes reduce $\mathcal{T}_{strong}$.
Among these types of attacks, the threat becomes significantly stronger if an attacker can remotely collect multiple timing traces of the same software or binary code. 
Let us assume that: (1) $T_i$ is the measured execution time for the $i^{th}$ run; (2) $n$ is the number of traces; (3) $\mu_0$, $\mu_1$ are the mean timings for different secrets (e.g., key bit 0 vs key bit 1); (4) $\sigma$ is the standard deviation of the noise in timing. Additionally, the signal-to-noise ratio (SNR) improves with more traces. Fig.~\ref{fig:threat} shows an overview of the discussed threat model.
% \vspace{-0.1in}
\begin{equation}
    SNR =  \frac{|\mu_0-\mu_1|}{\sigma /\sqrt{n}}
\end{equation}
% Tasneem: This means that increasing the number of traces ($n$) enhances the ability to distinguish between secret-dependent timing differences, making the attack more powerful even with remote access.
%rewrite by Tanzim

% This implies that as the number of traces (n) increases, the attacker’s ability to detect secret-dependent timing variations improves, thereby making the attack more effective and stronger, even in remote access scenarios.

% \vspace{-0.2in}
\subsection{Vulnerability Analysis}
Branch on Secret (BoS) and Loop on Secret (LoS) are the two dominant ways to capture secret data influence program runtime through control flow. HASTE models timing leakage by identifying secret-dependent branches and secret-dependent loop bounds as key sources of measurable runtime variation. Many practical embedded timing leaks reduce to these patterns, arising from secret-dependent paths or loop counts.

\subsubsection{Branch on Secret Analysis}
\begin{figure}[ht]
\centering

\includegraphics[width=\linewidth]{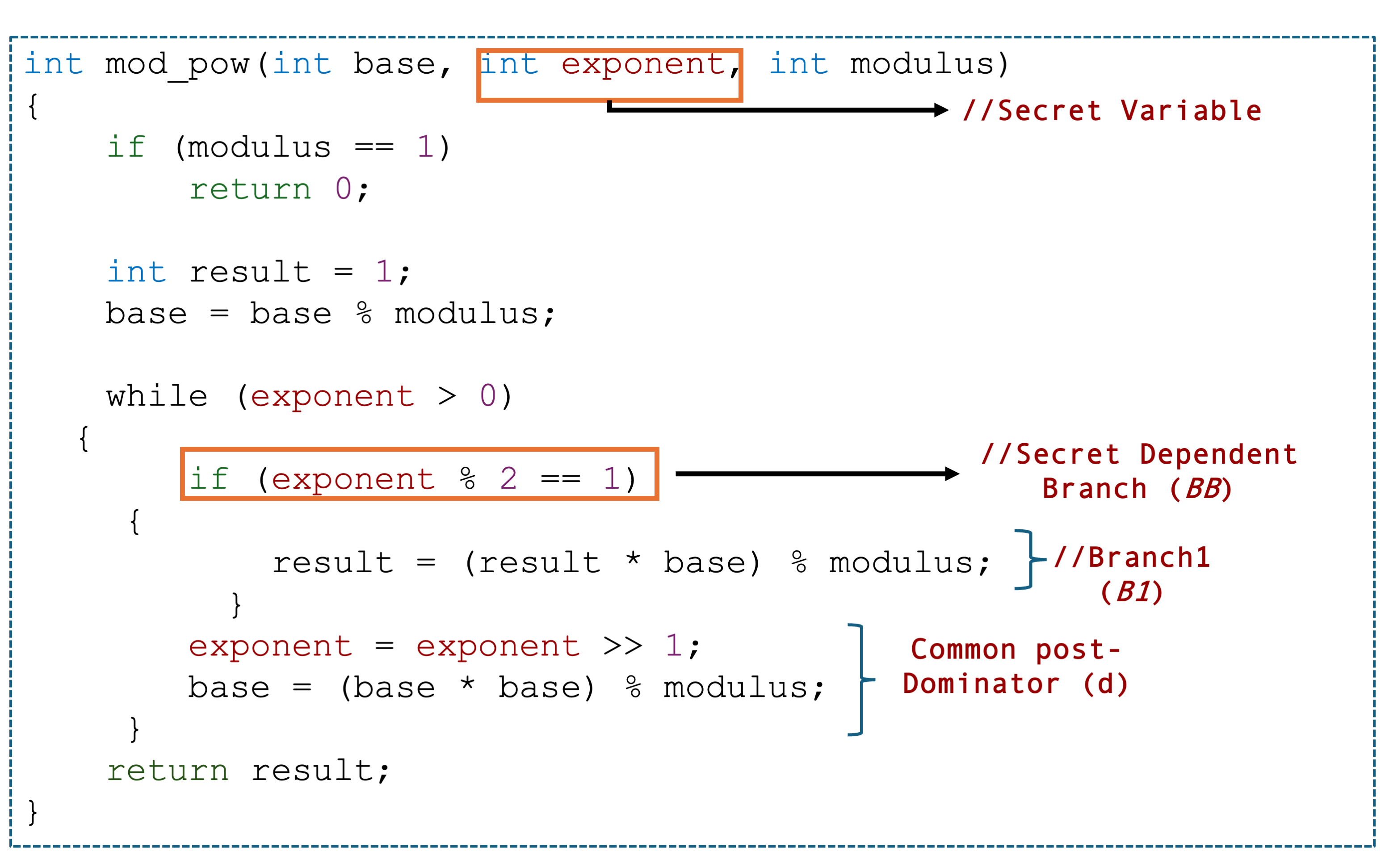}
\caption{\small{Branch on Secret (BoS) vulnerability.}\label{bos-exp}}

% \vspace{-0.2in}
\end{figure}

A Branch-on-Secret (BoS) vulnerability occurs when a secret-dependent condition controls program flow. BoS severity is an estimate of the maximum timing difference introduced by a branch whose direction depends on a secret variable~\cite{chakraborty2021haste}. If an attacker’s timing measurement precision (sensitivity) is finer than this value, the system is considered vulnerable. In practical terms, if an attacker can resolve timing differences smaller than the BoS severity, the secret-dependent branch becomes exploitable.
Let $B1$ and $B2$ be the two branch successors of a basic block $BB$ that depends on a secret variable where $d$ is the first common post-dominator of $B1$ and $B2$. Let $T()$ represent the execution time of a path. \textcolor{black}{A simplified BoS\_Severity is:}
 \begin{equation}
     BoS\_Severity = |T(B1\rightarrow{d}) - T(B2\rightarrow{d})|
 \end{equation}
The higher this difference, the easier it is for an attacker to distinguish which branch was taken on the basis of timing analysis. The smaller that gap, the harder it is for an attacker to tell which path ran. For example, in Fig.~\ref{bos-exp}, the secret is the value of $exponent$. Each loop iteration tests the exponent’s lowest bit with \texttt{if (exponent \% 2 == 1)}. If set (path B1), it performs an extra modular multiply, $result=(result*base)\%modulus$; otherwise (path B2), it skips this step. Both paths rejoin at $exponent >> 1$. 
% Because one path includes a big multiply and the other does not, the two paths take very different amounts of time. Anyone who can measure time more accurately than that or average many runs, can tell which path was taken and learn the exponent bit by bit.
%\centering

\subsubsection{Loop on Secret Analysis}
\begin{figure}[ht]
\centering

\includegraphics[width=\linewidth]{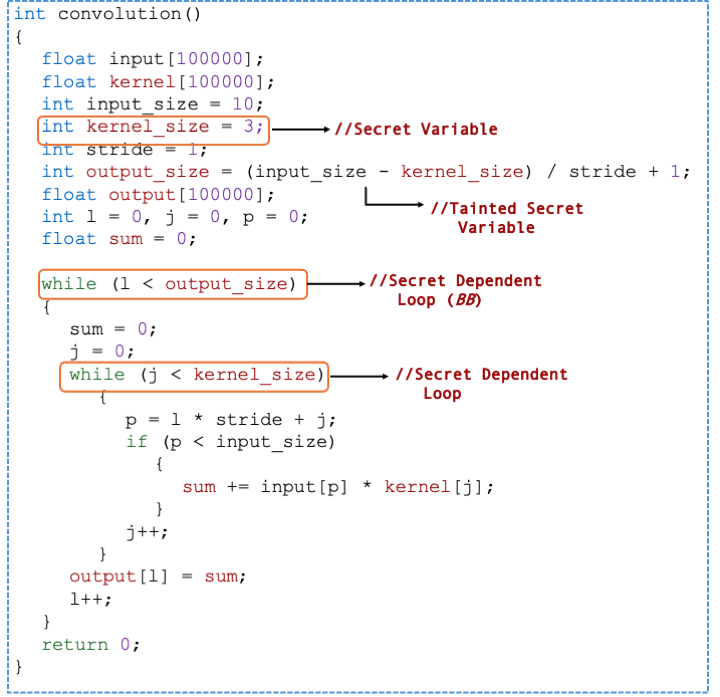}
\caption{\small{Loop on Secret (LoS) vulnerability.}\label{los-exp}}

\vspace{-0.2in}
\end{figure}

When the number of iterations in a loop is controlled by the secret is called loop-on-secret~\cite{chakraborty2021haste}. The LoS\_Resilience is the minimum number of loop iterations (controlled by a secret) required for a timing difference to become discernible to an attacker. Let $BB$  be the loop head (secret-dependent) and $\Delta t$ be the per-iteration timing difference. Let $\mathcal{T}$ be the smallest timing difference an attacker can reliably measure.
So, the LoS\_Resilience will be:
\begin{equation}
    LoS\_Resilience = |\frac{\mathcal{T}}{\Delta t}|
\end{equation}
% \vspace{-0.2in}
\begin{equation}
    LoS\_Resilience= \min_{n \in \mathbb{N}}\bigg\{n|n \times \Delta t \ge \mathcal{T}\bigg\}
\end{equation}
Here, $\Delta t$ is the per iteration time difference derived from the hardware-level basic block timings in the control flow graph. \textit{n} is the number of loop iterations controlled by the secret. 
In other words, the loop must run at least LoS\_Resilience times before the attacker can reliably discern a timing difference attributable to the secret-controlled loop count. So, if the number is large, the loop has to run a long time before any timing difference becomes visible, making the leak impractical and if the value is small even a tiny change in the secret produces a measurable time shift, so the code will be highly vulnerable. For example, in Fig.~\ref{los-exp}, the secret \texttt{kernel\_size} controls both the inner-loop bound, \texttt{while (j < kernel\_size)}, and, through \texttt{output\_size}, the outer-loop bound, \texttt{while (l < output\_size)}. Each inner-loop iteration performs one multiply--accumulate with nearly fixed cost $\Delta t$, so one outer-loop iteration costs about $\texttt{kernel\_size}\cdot\Delta t$. Since the outer loop repeats this work \texttt{output\_size} times, total runtime remains a deterministic function of \texttt{kernel\_size}; an attacker with timing resolution $T$ can infer the secret when $\texttt{kernel\_size}\cdot\Delta t \ge T$.
% The smallest $kernel\_size$ that satisfies this inequality is the LoS Resilience $T⁄\Delta t$.
% if the secret makes the loops run at least this many extra iterations, the timing difference becomes observable.
%\centering

\subsection{Related Work}
\begin{table*}[!htbp]
\captionsetup{
  justification=centering,
  % labelfont={color=blue},
  % textfont={color=blue}
}
\centering
\caption{Comparing \name~with existing software timing side-channel analysis/mitigation approaches.}
\label{tab:comparison}

 % make all table text blue
\arrayrulecolor{black} % keep borders black

\renewcommand{\arraystretch}{1.12}
\small\addtolength{\tabcolsep}{-2.5pt}
\begin{tabular}{|
>{\centering\arraybackslash}p{0.109\textwidth}|
>{\centering\arraybackslash}p{0.19\textwidth}|
>{\centering\arraybackslash}p{0.109\textwidth}|
>{\centering\arraybackslash}p{0.09\textwidth}|
>{\centering\arraybackslash}p{0.095\textwidth}|
>{\centering\arraybackslash}p{0.31\textwidth}|}
\hline
\rowcolor[HTML]{ECF4FF} 
\textbf{Study} &
  \textbf{Framework Objective} &
  \textbf{Analysis Type} &
  \textbf{Repair Scope} &
  \textbf{Hardware Aware} &
  \textbf{Repair Mechanism} \\ \hline

\rowcolor[HTML]{EFEFEF} 
\textbf{PENDULUM\textsuperscript{*}\cite{pendulum}} &
  Identify timing vulnerabilities and constant-time fixing &
  Dynamic &
  Java &
  N/A &
  Pattern based source repair using conditional assignments and loop bound rewriting \\ \hline

\textbf{DifFuzzAR\textsuperscript{*}\cite{diffuzzar}} &
  Identify timing vulnerabilities and constant-time fixing &
  Dynamic &
  Java &
  N/A &
  Source refactoring using single exit conversion and mirrored branch computation \\ \hline

\rowcolor[HTML]{EFEFEF} 
\textbf{CT-Wasm\cite{watt2019ct}} &
  Identify vulnerabilities and enforce constant-time &
  Formal &
  Assembly &
  N/A &
Type system rewrite in CT-Wasm; no post-hoc repair \\ \hline

\textbf{dudect\cite{reparaz2017dude}} &
  Detect non-constant time program &
  Dynamic &
  N/A &
  Architecture &
 Detection only via statistical timing tests \\ \hline

\rowcolor[HTML]{EFEFEF} 
\textbf{FlowTracker\cite{rodrigues2016sparse}} &
  Detect timing-dependent control flow and data access patterns &
  Static &
  N/A &
  N/A &
  Detection only via static implicit flow analysis \\ \hline

\textbf{Nemisis\cite{salehi2022nemesisguard}} &
Balance secret-dependent branch, mitigate nemesis attack &
  Dynamic &
  Binary &
  Partial &
Binary rewriting to balance secret-dependent branches \\ \hline

\rowcolor[HTML]{EFEFEF} 
\textbf{HASTE\cite{chakraborty2021haste}} &
Quantify and localize timing side-channel vulnerabilities &
  Static+Dynamic &
  N/A &
  Simulation &
Detection/quantification only using hardware-aware CFG analysis \\ \hline

\rowcolor[HTML]{f1fef9} 
\textbf{\name~(ours)} &
  Quantify, localize, and repair timing side channel vulnerabilities &
  Static+Dynamic &
  C/C++/Java &
  Real Hardware &
Hardware-aware padding to satisfy BoS/LoS thresholds on target hardware \\ \hline

\bottomrule
\end{tabular}

\raggedright\footnotesize{\textit{*: State of the Art (SOTA)}}
 % end blue text group

\vspace{-0.2in}
\end{table*}
FlowTracker~\cite{rodrigues2016sparse} is a static information-flow analysis tool for detecting timing leaks. Rodrigues et al. introduce a sparse representation of implicit flows that computes control dependencies in a single dominance-tree traversal, yielding a number of edges linear in program size. It only flags vulnerable branches and cannot repair code or measure real hardware timing. Dudect~\cite{reparaz2017dude} takes a complementary, dynamic approach. It is a lightweight black-box testing tool that runs code on target hardware and uses statistical hypothesis testing to check if execution time differs for different secret inputs. The advantage of Dudect is practicality as it reflects actual hardware behavior and is easy to deploy. However, it only answers `constant-time or not' for a given routine and it does not localize the leak or suggest fixes. CT-Wasm~\cite{watt2019ct} represents a language-based defense where Watt et al. define Constant-Time WebAssembly (CT-Wasm), a strictly typed extension of Wasm that enforces that well-typed programs cannot leak secret data via timing. By building a verified, sound type checker, authors even supply a compiler pass to rewrite existing crypto libraries into CT-Wasm; for example, a TweetNaCl implementation compiled under CT-Wasm is shown to run in equal time for all inputs. This formal approach is powerful because it provides provable guarantees of no timing leakage. Potential limitations are: it applies only to code written or ported to the CT-Wasm subset and it is a static method that requires developers to use a specialized language/engine (e.g., browser or Node with CT-Wasm support). NemesisGuard~\cite{salehi2022nemesisguard} targets microarchitectural `Nemesis' side channels from interrupt latency in trusted execution environments (TEEs). It is a static binary-instrumentation framework that pads secret-dependent code paths to equalize the cost of handling an interrupt. However, it has no focus on addressing data-dependent timing.

DifFuzzAR~\cite{diffuzzar} is an automated repair tool, focused on Java, which applies fixes after DifFuzz~\cite{dfzold} finds a vulnerable method. It is especially effective on control-flow-based leaks, which makes it a suitable platform to compare with our \name~framework. DifFuzzAR's goal is to produce functionally correct, constant-time code. PENDULUM~\cite{pendulum} is another recent automated repair tool for timing leaks at the source level (Java). Ruan et al. describe PENDULUM as the first approach that automatically locates and fixes timing side-channel vulnerabilities in code. It uses fuzzing to quantitatively estimate the amount of leakage in different program regions, then applies pattern-driven source transformations to equalize execution time, outperforming DifFuzzAR. However, the repairs come at a cost; the patches incur substantial slowdowns compared to the original code. PENDULUM represents a state-of-the-art source-level mitigation, but its limitations include high fixed-code overhead and the lack of on-device validation. In contrast, \name~combines software analysis with hardware-in-the-loop measurement to assess and mitigate timing leaks on real embedded platforms, complementing and extending the capabilities of these existing approaches at low overhead.

The repair mechanisms of these tools are also different. Unlike PENDULUM and DifFuzzAR, which mitigate timing leakage through device-agnostic source-to-source transformations, \name~does not enforce a single generic balancing template. DifFuzzAR repairs Java code by source-level refactoring, including removal of early exits, restructuring of secret-dependent conditionals by mirroring branch computations with fresh temporaries, and, when possible, replacement of secret-dependent loop guards with public arguments. PENDULUM performs quantification-guided, pattern-based repair by rewriting secret-dependent assignments into constant-time conditional assignments and transforming secret-dependent loops to execute up to a secret-independent upper bound while making extra iterations functionally inert. In contrast, \name~uses cycle measurements from the target device to insert only the amount of additional work needed to reduce the measured BoS and LoS leakage below a specified attacker threshold on the deployment hardware. In summary, in Table~\ref{tab:comparison}, we provide a qualitative comparison of \name~with existing frameworks designed for detecting and mitigating timing side-channel vulnerabilities.

\section{Motivation}
% Our research is based on the following motivations: need for hardware-awareness during fixing 
% Next we will discuss the two key questions that have motivated our research.

\subsection{Why Hardware-Awareness During Mitigation?}
The majority of existing mitigation approaches (discussed above) focus on addressing timing mismatches by enforcing complete control-flow balancing (constant-time programming \cite{10.1145/3613424.3623796, barthe2018secure}).
% Constant-time programming has long been the default solution for closing timing side-channel leaks by forcing every control path to take exactly the same number of steps so that developers can be sure no attacker will learn secrets timing observation. 
On many edge and IoT platforms the extra branches, arithmetic masks, and dummy memory accesses needed to equalize every path slow programs dramatically, swell code size, and consume precious power units. 
% Moreover, constant time codes are typically crafted and benchmarked on desktop CPUs, so their penalties can grow even larger once the same code is cross-compiled to low-power cores with shallow pipelines and small caches. 
These realities motivate our work \name~ rather than `one fixed solution' for all, we pair lightweight program analysis with hardware-in-the-loop timing measurements (dynamic across any hardware device) to identify only the code regions that truly leak on the actual device under the actual workload, and then apply the repairs balancing low overhead.

\subsection{Why Consider Specific Attack Scenario?}
As outlined in Section~\ref{threat model}, the feasibility and severity of a timing side‑channel attack depend critically on the attacker’s level of access to the target system. When an adversary can interact with the device only remotely every measurable delay must be large enough to survive network jitter and operating system noise. For example, this pushes the timing sensitivity threshold up to 5-50 CPU cycles. Subtle variations like 1-5 cycles are therefore unlikely to be distinguishable and the system appears secure. The situation changes markedly when the attacker has local or physical access (leading to near cycle accurate attack).  
% In this case, far less environmental noise is present, and timing differences is small (e.g., 8 CPU cycles) which can be resolved.  If the same implementation leaks 10 cycle differences, a local adversary can mount a reliable, strong attack especially if they are permitted to record multiple traces. 
Ignoring the attack scenario during mitigation risks both extremes: under‑fixing and over‑fixing.
% Hence effective countermeasures must be tailored to a realistically assumed adversary model.
% , balancing secrecy while provide preservation against area, latency, and execution‑time overheads.

% \input{motivation}

\section{\name~Methodology}
To address the shortcomings of state-of-the-art timing side-channel vulnerability mitigation frameworks, we develop \name~ that automatically: (1) Performs timing leakage evaluation directly on real IoT/Edge devices; (2) Optimally mitigates timing imbalances based on a given threat model and timing information from the system (hardware + software); (3) Supports the detection/fixing of \textit{Branch on Secret (BoS)} and \textit{Loop on Secret (LoS)} vulnerabilities by considering some important factors: (i) Branch prediction effect, (ii) Cache states effect. Fig.~\ref{fig:haste-fix} illustrates the complete workflow of \name.

% The precision with which an attacker can measure the execution time of a program depends on their access level. For example, a local attacker with physical access can measure time more accurately than a remote attacker. We take inspiration from \cite{chakraborty2021haste, 10256749} and propose a joint hardware-software strategy to mitigate timing attacks without excessive performance overhead.  a targeted repair approach. The details of TiLeR are described below.

\subsection{Hardware-Software CPU Cycles Extraction}
\begin{algorithm}[!ht]
\setstretch{1}
\caption{Extract CPU Cycles}
\label{algo: Cycle extraction}
\DontPrintSemicolon % Ensures semicolons are not printed at the end of lines
\KwIn{$[\mathcal{CFG}, \mathcal{N}_{Run}]$}
\KwOut{$[\mathcal{CFG}_{new}]$}
$\mathcal{CFG}_{new}\gets \emptyset$\;
$\mathbb{C} \gets []$\;
$\overline{\mathbb{C}} \gets 0$\;
$\mathcal{C}_{real} \gets 0$\;
$\mathbb{R} \gets 1000000$\;
\For {$\mathcal{BB}$ $\in$ $\mathcal{CFG}$ \tcp{$\mathcal{BB}$: Basic Block}}
{
    $envBB$ $\gets$ \textbf{\textit{EncapsulateBB}} ($\mathcal{BB})$\;
    $envBB.exe \gets \textbf{\textit{Compile}}(envBB)$\;

    \For {$i$ $\in$ $\mathcal{N}_{Run}$}
    {
        $\mathcal{C} \gets \textbf{\textit{CalculateCycle}}(EnvBB.exe)$\;
        $\mathbb{C}.append(\mathcal{C})$\;
    }
    $\overline{\mathbb{C}}  \gets \textbf{\textit{sum}}(\mathbb{C})/\mathcal{N}_{Run}$\;
    $\mathbb{C}_{real} \gets \overline{\mathbb{C}} /\mathbb{R}$\;
    $\mathcal{CFG}_{new}[\mathcal{BB}][``cpuCycle"] \gets \mathcal{C}_{real}$\;
}
\Return $\mathcal{CFG}_{new}$\;
% \vspace{-0.2in}
\end{algorithm}
% \vspace{-0.2in}
To extract CPU cycles for a given software code, a control flow graph (CFG) is generated by creating adjacency relationships for each basic block based on program statements, branching, and loops. CPU cycle count for each basic block, based on the hardware architecture, is determined. Algorithm~\ref{algo: Cycle extraction} automates the retrieval of CPU cycles for each basic block, using inputs: (i) $\mathcal{CFG}$, the control flow graph of the benchmarked software; (ii) $\mathcal{N}_{Run}$, specifying how many times to measure cycles for each block.
We first iterate over all basic blocks ($\mathcal{BB}$) to encapsulate them within a $for\_loop$ set to $\mathbb{R}$ = 1,000,000 iterations, enabling precise cycle measurement through the $EncapsulateBB()$ function. This function, (i) initializes the blocks' live variables; (ii) prepares the cache state. The encapsulated blocks are then compiled into an executable in lines 6-9. To capture cache sensitivity without contaminating the block logic, $EncapsulateBB()$ performs randomized cache warming immediately before timing. On each run, it draws a seed and a warm fraction (both overridable via environment knobs), selects that proportion of the block’s live variables, and touches one cache line per selected variable. The encapsulated blocks are compiled into an executable using compile-time instrumentation, which modifies the program before execution.  
In lines 10-13, CPU-cycles ($\mathcal{C}_{real}$) for each block are measured $\mathcal{N}_{Run}$ times, with the average cycles ($\overline{\mathbb{C}}$) computed to capture the variable timing effect on the CPU-cycles. We consider random input values for each of the given benchmarks to evaluate the CPU-cycle because basic-block execution time can vary with input-dependent data values, memory-access behavior, cache state, and branch behavior. Using random inputs across repeated runs helps capture representative timing variation and prevents the cycle estimate from being biased toward a single fixed input instance. To determine the $\mathcal{C}_{real}$, we divide the $\overline{\mathbb{C}}$ by the $for\_loop$ iteration count $\mathbb{R}$. Finally, $\mathcal{CFG}$ is updated ($\mathcal{CFG}_{new}$) with the measured CPU cycles for each basic block.

\subsection {Taint Analysis}
\begin{algorithm}[!ht]
\setstretch{1}
\caption{Construct Candidate Sensitive Set}
\label{algo:taint_analysis}
\DontPrintSemicolon
\KwIn{[${Prg}, \mathcal{S}_{seed}]$}
\KwOut{$[\mathcal{S}_{taint}, \mathcal{S}_{ctrl}]$}

$\mathcal{S}_{taint} \gets \mathcal{S}_{seed}$\;
$\mathcal{S}_{ctrl} \gets \emptyset$\;
$\mathcal{W} \gets \mathcal{S}_{seed}$\;
$\mathcal{G} \gets \textbf{\textit{BuildProgramGraph}}(Prg)$\;

\While{$\mathcal{W} \neq \emptyset$}
{
    $v \gets \textbf{\textit{Pop}}(\mathcal{W})$\;

    \For{$stmt \in \textbf{\textit{UseStmts}}(\mathcal{G}, v)$}
    {
        \If{$stmt$ defines $x$ from $v$}
        {
            \If{$x \notin \mathcal{S}_{taint}$}
            {
                $\mathcal{S}_{taint} \gets \mathcal{S}_{taint} \cup \{x\}$\;
                $\mathcal{W}.append(x)$\;
            }
        }

        \If{$v$ flows to a parameter or return value in $stmt$}
        {
            $u \gets \textbf{\textit{ResolveFlow}}(stmt,v)$\;
            
            \If{$u \notin \mathcal{S}_{taint}$}
            {
                $\mathcal{S}_{taint} \gets \mathcal{S}_{taint} \cup \{u\}$\;
                $\mathcal{W}.append(u)$\;
            }
        }

        \If{$v$ influences a branch predicate or loop bound in $stmt$}
        {
            $\mathcal{S}_{ctrl} \gets \mathcal{S}_{ctrl} \cup \{v\}$\;
        }
    }
}

\Return $[\mathcal{S}_{taint}, \mathcal{S}_{ctrl}]$\;
\end{algorithm}

Algorithm \ref{algo:taint_analysis} takes the program ${Prg}$ and the user-defined sensitive seed set $\mathcal{S}_{seed}$ as inputs and returns $\mathcal{S}_{taint}$, the full set of variables derived from sensitive seeds, and $\mathcal{S}_{ctrl}$, the subset that affects control flow and is passed to the BoS/LoS analysis stage. First, it initializes the tainted set $\mathcal{S}_{taint}$ defined by the user and creates an empty sensitive-control set $\mathcal{S}_{ctrl}$. Next, it places all seed variables into a worklist $\mathcal{W}$, which stores variables whose data flow still needs to be analyzed. Then it builds a program graph $\mathcal{G}$ from the source code to capture assignments, variable uses, function calls, return values, branches, and loops (line 4). While the worklist is not empty, it removes one variable $v$ and checks every statement where $v$ is used (lines 6,7). From lines 8-11, if a statement defines a new variable $x$ from $v$, then $x$ is added to the tainted set and inserted into the worklist for further propagation. If $v$ flows through a function parameter or return value, it resolves the corresponding variable $u$, adds it to the tainted set, and continues propagation from $u$ (lines 12--16). Finally, if $v$ appears in a branch predicate or loop bound, it is added to $\mathcal{S}_{ctrl}$ because it may create a BoS or LoS vulnerability from line 17 to 18.
% \vspace{-0.25 in}
\subsection{Branch on Secret Severity Mitigation}
\begin{algorithm}[ht]
\setstretch{1}
\caption{\name~ for BoS}
\label{algo: HASTE-FIX}
\DontPrintSemicolon % Ensures semicolons are not printed at the end of lines
\KwIn{$[benchmark, \mathcal{S}, \mathcal{T}, \mathcal{N}_{Run},, \mathcal{B}_{t1}, \mathcal{B}_{t2}, \mathcal{P}]$}
\KwOut{$[fixCode]$}
$G \gets \emptyset$\;
$\mathcal{CFG} \gets \textbf{\textit{GetTaintedCFG}}(benchmark, \mathcal{S})$\;
$\mathcal{CFG} \gets \textbf{\textit{GetCycle}}
(\mathcal{CFG},\mathcal{N}_{Run})$\;
$\mathcal{CFG} \gets \textbf{\textit{Splice}}(\mathcal{CFG})$\;
$\mathcal{CFG} \gets \textbf{\textit{AssignBranchTimes}}(\mathcal{CFG},\ \mathcal{B}_{t1},\ \mathcal{B}_{t2},\ \mathcal{P})$ \tcp*{push blended times onto spliced edges: $t_{\text{long}}=\mathcal{B}_{t1}\mathcal{P}+\mathcal{B}_{t2}(1-\mathcal{P})$, $t_{\text{short}}=\mathcal{B}_{t1}(1-\mathcal{P})+\mathcal{B}_{t2}\mathcal{P}$}

\While {$True$}
{
    \For {$\mathcal{BB} \in \mathcal{CFG}$ \tcp{$\mathcal{BB}$: Basic Block}} % BB: Basic Block
    {
        \If {$\textbf{\textit{BranchOnSecret}}(\mathcal{BB}, \mathcal{S})$}
        {
            $postDom = \textbf{\textit{GetPostDoms}}(\mathcal{BB}, \mathcal{CFG})$ \;
            $\mathbb{B}_1 , \mathbb{B}_2$  = \text{Immediate BB successors} \;
            $uniqDom \gets \textbf{\textit{FindUniqDom}}(postDom)$\;
            $\mathbb{L}_1 \gets \textbf{\textit{FindLongestPath}}(\mathbb{B}_1, uniqDom)$\;
            $\mathbb{L}_2 \gets \textbf{\textit{FindLongestPath}}(\mathbb{B}_2, uniqDom)$\;
            $\mathbb{S}_1 \gets \textbf{\textit{FindShortestPath}}(\mathbb{B}_1, uniqDom)$\;
            $\mathbb{S}_2 \gets \textbf{\textit{FindShortestPath}}(\mathbb{B}_2, uniqDom)$\;

            \If{$|\mathbb{L}_1-\mathbb{S}_2|>\mathcal{T} \lor |\mathbb{L}_2-\mathbb{S}_1|>\mathcal{T}$ }
            {   \If{$|\mathbb{L}_1-\mathbb{S}_2|>|\mathbb{L}_2-\mathbb{S}_1|$ }
                    {
                    $\mathcal{P}_{shortest} \gets \mathbb{S}_2$\;
                    $G \gets \textbf{\textit{AddNoise}}(\mathbb{S}_2, \mathcal{CFG})$\;
                    $\mathcal{CFG} \gets \textbf{\textit{GetCycle}}(G, \mathcal{N}_{Run})$\;
                    }
                \Else 
                {
                $\mathcal{P}_{shortest}  \gets \mathbb{S}_1$\;
                $G \gets \textbf{\textit{AddNoise}}(\mathbb{S}_1, \mathcal{CFG})$\;
                    $\mathcal{CFG} \gets \textbf{\textit{GetCycle}}(G, \mathcal{N}_{Run})$\;
                    }
                }
                $\textbf{continue}$\; 
            }    
            \Else
            {
                $fixCode \gets \textbf{\textit{ConvCFG2Code}}(\mathcal{CFG})$\;
            }   
            }
        }
\Return{$fixCode$}
\end{algorithm}
% \vspace{-0.2in}

To effectively mitigate the BoS vulnerability within the software code while considering the attacker's timing sensitivity, we introduce Algorithm~\ref{algo: HASTE-FIX}. This algorithm offers a targeted approach to repair vulnerability (BoS) using the corresponding hardware architecture, the algorithm takes inputs as:
(i) $benchmark$ is the source code to be analyzed; (ii) $\mathcal{S}$, the filtered sensitive-control set containing variables generated from Algorithm~\ref{algo:taint_analysis} (iii) $\mathcal{T}$ is the attacker's timing sensitivity, which is a measure of the attacker's ability to exploit timing side-channel attacks; (iv) $\mathcal{N}_{Run}$, specifying how many times to measure cycles for each block; (v) $\mathcal{B}_{t1}$ and $\mathcal{B}_{t2}$ are the two timing values for two different branches to consider the branching effect; and (vi) $\mathcal{P}$ is the probability of selecting one particular branch.The algorithm then generates a tainted $\mathcal{CFG}$ using the $GetTaintedCFG()$ function. After that using the $GetCycle()$ function (based on Algorithm~\ref{algo: Cycle extraction}) calculates the CPU cycles and updates the $\mathcal{CFG}$ (line 2-3). From lines 4-5, $Splice()$ function splices each two-way branch by inserting synthetic edge nodes, one per successor path. Splicing lets us assign per successor timing without mutating basic blocks. For every secret-dependent branch, it locates the post dominating join of its two successors and computes the longest/shortest path costs from each successor path to the join using Bellman–Ford relaxations, then pushes edge times onto the spliced nodes to model the indirect branching effect using user provided inputs $\mathcal{B}_{t1}$, $\mathcal{B}_{t2}$ and $\mathcal{P}$. In lines 6-10, the algorithm iteratively traverses the $\mathcal{CFG}$, identifying basic blocks ($\mathcal{BB}$) whose branching decisions are influenced by $\mathcal{S}$. For each such block, it computes the post-dominator set ($postDom$) using the $GetPostDoms()$ function. By analyzing the longest and shortest paths to the unique immediate post-dominator (lines 11-12) when the Branch-on-Secret (BoS) path-length discrepancy exceeds the attacker’s timing sensitivity \(\mathcal{T}\), compile-time instrumentation (lines 13–25) invokes the $AddNoise()$ function to inject non-operational (dummy) instructions into the shorter execution  path to reduce the timing imbalance between secret dependent paths. To prevent compiler optimization from removing the padding, it does not use empty loops or unused arithmetic instructions, instead, it generates semantically neutral operations that update ``volatile" variables. Since these updates are treated as observable side effects, the compiler preserves the inserted operations while the original program output remains unchanged. This process is repeated until the BoS severity falls below the $\mathcal{T}$ threshold. The modified $\mathcal{CFG}$ is then translated back to software code (lines 26-28).

\begin{table*}[]
\centering
\caption{Benchmarks for evaluating \name~ with constant-time programming, DifFuzzAR \cite{diffuzzar}, and PENDULUM \cite{pendulum}.}
\label{tab:benchmark_deatils}
\renewcommand{\arraystretch}{1}
\small\addtolength{\tabcolsep}{6pt}
\label{tab:my-table}
% \resizebox{\textwidth}{!}{%
\begin{tabular}{|c|l|l|c|}
\hline
\rowcolor[HTML]{ECF4FF} 
\textbf{Language} &
  \multicolumn{1}{c|}{\cellcolor[HTML]{ECF4FF}\textbf{Benchmarks}} &
  \multicolumn{1}{c|}{\cellcolor[HTML]{ECF4FF}\textbf{Description}} &
  \textbf{Vulnerabilities} \\ \hline
 &
  blazer\_modpow1 &
  modPow1\_unsafe function &
  BoS, LoS \\ \cline{2-4} 
 &
  blazer\_array &
  array\_unsafe function &
  BoS \\ \cline{2-4} 
 &
  blazer\_sanity &
  sanity\_unsafe &
  BoS \\ \cline{2-4} 
 &
  blazer\_straightline &
  straightline\_unsafe &
  BoS \\ \cline{2-4} 
 &
  blazer\_unixlogin &
  login\_unsafe &
  BoS, LoS \\ \cline{2-4} 
 &
  themis\_boot-stateless-auth &
  unsafe\_isEqual &
  BoS, LoS \\ \cline{2-4} 
 &
  themis\_picketbox &
  validatePassword\_unsafe &
  BoS, LoS \\ \cline{2-4} 
 &
  blazer\_passwordEq &
  passwordsEqual\_unsafe &
  BoS, LoS \\ \cline{2-4} 
 &
  example\_PWCheck &
  pwcheck1\_unsafe &
  BoS, LoS \\ \cline{2-4} 
\multirow{-10}{*}{\textbf{Java\textsuperscript{*}} } &
  themis\_jdk &
  isEqual\_unsafe &
  BoS, LoS \\ \hline
 &
  Benchmark 1 &
  RSA Mod Pow Exponentiation Function~\cite{6021216} &
  BoS, LoS \\ \cline{2-4} 
 &
  Benchmark 2 &
  Relu Activation Function~\cite{ramachandran2017searching} &
  BoS, LoS \\ \cline{2-4} 
 &
  Benchmark 3 &
  Leaky Relu Activation Function~\cite{Leaky-relu} &
  BoS, LoS \\ \cline{2-4} 
 &
  Benchmark 4 &
  Sigmoid Activation Function~\cite{Sigmoid} &
  BoS, LoS \\ \cline{2-4} 
 &
  Benchmark 5 &
  Data Compression (Run Length Encoding)~\cite{DataCompression} &
  BoS, LoS \\ \cline{2-4} 
 &
  Benchmark 6 &
  CNN (Convolution Operation) Function ~\cite{Convolution} &
  BoS, LoS \\ \cline{2-4} 
 &
  Benchmark 7 &
  ElGamal Mod Pow Exponentiation Function~\cite{elgamal1985public} &
  BoS, LoS \\ \cline{2-4} 
 &
  Benchmark 8 &
  Diffie Hellman Mod Pow Exponentiation Function~\cite{diffie2022new} &
  BoS, LoS \\ \cline{2-4} 
 &
  Benchmark 9 &
  Feistal Cipher (Round Function)~\cite{feistel1973cryptography} &
  BoS, LoS \\ \cline{2-4} 
 &
  Benchmark 10 &
  XOR Stream Cipher (Bitwise Exclusive‑OR Operation)~\cite{schneier2007applied} &
  BoS \\ \cline{2-4} 
 &
  Benchmark 11 &
  Hill Cipher (Matrix‑multiplication‑mod‑26 Function)~\cite{hill1929cryptography} &
  BoS \\ \cline{2-4} 
\multirow{-12}{*}{\textbf{C++/C}} &
  Benchmark 12 &
  RC4 Cipher (Key‑Scheduling  (KSA) Function)~\cite{paul2011rc4} &
  BoS \\ \hline
\bottomrule
\end{tabular}%
% } // ResizeBox
\\
\raggedright\footnotesize{\textit{*All the Java codes (fixed) are taken from \cite{pendgit} \cite{dfzargit} }}

% \vspace{-0.2in}

\end{table*}
% \usepackage{graphicx}
\begin{table}[ht]
\centering
\caption{Hardware devices used in our experiments.}
\label{tab:my-devices}
\renewcommand{\arraystretch}{1}
\small\addtolength{\tabcolsep}{-1pt}
\begin{tabular}{|l|c|}
\hline
\rowcolor[HTML]{DAE8FC} 
\multicolumn{1}{|c|}{\cellcolor[HTML]{DAE8FC}\textbf{Devices}} & \multicolumn{1}{c|}{\cellcolor[HTML]{DAE8FC}\textbf{Description}} \\ \hline
Devices 1 & {\color[HTML]{000000} 11th Generation Intel Core i9-11900H Processor} \\ \hline
\rowcolor[HTML]{EFEFEF} 
Devices 2 & Jetson Nano ARM Cortex-A57 MPCore processor                    \\ \hline
Devices 3 & Jetson Orin Nano ARM Cortex-A78AE processor                    \\ \hline
\rowcolor[HTML]{EFEFEF} 
Devices 4 & Jetson AGX Xavier Carmel ARM v8.2                              \\ \hline

Devices 5 & Jetson Xavier NX Carmel ARM v8.2                              \\ \hline
\rowcolor[HTML]{EFEFEF} 
Devices 6 & Raspberry Pi  Quad-core Cortex-A72                             \\ \hline
\end{tabular}
\vspace{-0.2in}
\end{table}

\subsection{Loop on Secret Severity Mitigation}
\begin{algorithm}[ht]
\setstretch{1}
\caption{\name~ for LoS}
\label{algo: HASTE-FIX-Los}
\DontPrintSemicolon % Ensures semicolons are not printed at the end of lines
\KwIn{$[benchmark, \mathcal{S}, \mathcal{T}, \mathcal{N}_{Run}, \mathcal{L}_{min}, , \mathcal{B}_{t1}, \mathcal{B}_{t2}, \mathcal{P}]$}
\KwOut{$[fixCode]$}
$G \gets \emptyset$\;
$\mathcal{CFG} \gets \textbf{\textit{GetTaintedCFG}}(benchmark, \mathcal{S})$\;
$\mathcal{CFG} \gets \textbf{\textit{GetCycle}}(\mathcal{CFG},\mathcal{N}_{Run})$\;
$\mathcal{CFG} \gets \textbf{\textit{Splice}}(\mathcal{CFG})$\;
$\mathcal{CFG} \gets \textbf{\textit{AssignBranchTimes}}(\mathcal{CFG},\ \mathcal{B}_{t1},\ \mathcal{B}_{t2},\ \mathcal{P})$ \tcp*{push blended times onto spliced edges: $t_{\text{long}}=\mathcal{B}_{t1}\mathcal{P}+\mathcal{B}_{t2}(1-\mathcal{P})$, $t_{\text{short}}=\mathcal{B}_{t1}(1-\mathcal{P})+\mathcal{B}_{t2}\mathcal{P}$}
\While {$True$}
{
    \For {$\mathcal{BB} \in CFG$ \tcp{$\mathcal{BB}$: Basic Block} } % BB: Basic Block
    {
        \If{$\textbf{\textit{LoopHeadOnSecret}}(\mathcal{BB},\mathcal{S})$}
        {
        $succBB=\text{Immediate BB successors} $\;
        $\mathbb{L} = \textbf{\textit{FindLongestPath}}(succBB, \mathcal{BB})$\;
        $\mathbb{S} = \textbf{\textit{FindLongestPath}}(succBB, \mathcal{BB})$\;
            \If{$\mathcal{T}/|\mathbb{L}-\mathbb{S}| < \mathcal{L}_{min}$}
            {
            $G \gets \textbf{\textit{AddNoise}}(\mathcal{S},\mathcal{CFG})$\;
            $\mathcal{CFG} \gets \textbf{\textit{GetCycle}}(G, \mathcal{N}_{Run})$\;
            $\textbf{continue}$\;
            }
            \Else
            {
                $fixCode \gets \textbf{\textit{ConvCFG2Code}}(\mathcal{CFG})$\;
            }
        }
        }
        }
\Return{$fixCode$}
\end{algorithm}
% \vspace{-0.2in}

We reduce the Loop-on-secret vulnerability using Algorithm~\ref{algo: HASTE-FIX-Los}. The algorithm takes inputs as (i) $benchmark$ is the source code to be analyzed; (ii) $\mathcal{S}$, the filtered sensitive-control set containing variables generated from Algorithm~\ref{algo:taint_analysis} (iii) $\mathcal{T}$ is the attacker's timing sensitivity which measure of the attacker's ability to exploit timing side-channel attacks; (iv) $\mathcal{L}_{min}$, a threshold for the minimum LoS Resilience; (v) $\mathcal{N}_{Run}$, specifying how many times to measure the cycles for each block; (vi) $\mathcal{B}_{t1}$ and $\mathcal{B}_{t2}$ are the two timing values for two different branches to consider the branching effect; and (vii) $\mathcal{P}$ is the probability of selecting a particular branch. 
Algorithm~\ref{algo: HASTE-FIX-Los} targets Loop-based side-channel (LoS) attacks. Identifies loops whose iteration count is influenced by secret variables using the $LoopHeadOnSecret()$ function on line 9. After finding the loop head on a secret, by calculating the longest and shortest paths within the loop body (lines 9-12), the algorithm assesses the loop's LoS resilience ($\mathcal{T}/|\mathbb{L}-\mathbb{S}|)$. If resilience falls below the minimum LoS Resilience ($\mathcal{L}_{min}$), compile-time instrumentation (lines 13–15) uses the $AddNoise()$ function to insert non-operational (dummy) instructions into the shorter path, thus increasing timing variability. The padding code is written using volatile updates, so the compiler treats the inserted operations as observable and does not remove them during optimization. This avoids unreliable padding patterns such as empty loops or unused computations. The modified $\mathcal{CFG}$ is then translated back into the software code.

\subsection{Modular Design for Scalability/Extensibility}
We present a language‑agnostic framework that identifies and mitigates software runtime vulnerabilities by recording the CPU cycle of every basic block in the program's control flow graph (CFG) to locate blocks whose timing breaches the adversarial threat model and hardware‑aware patching, which means inserting calibrated no‑operation (NOP) instructions that equalize the latencies of those blocks while considering microarchitectural constraints of the target platform. To demonstrate the robustness of the framework, we evaluated it on the C/C++ and Java environments. Although these languages differ in syntax, compilation model, and source file organization, the core workflow, cycle accurate severity analysis followed by architecture-conscious padding, remains identical across them.

% \usepackage{multirow}
\begin{table*}[!htbp]
\captionsetup{justification=centering}
\caption{Comparing the code-fixing/mitigation efficacy of ~\name~ with SOTA techniques for different edge devices and Java benchmarks. Here timing sensitivity $\mathcal{T} = 10$, PDL = PENDULUM \cite{pendulum}, and DFZ = DifFuzzAR \cite{diffuzzar}.  }
\label{tab:comparisonJava}
\renewcommand{\arraystretch}{1.15}
\small\addtolength{\tabcolsep}{-4.8pt}
\begin{tabular}{|c|l|ccc
>{\columncolor[HTML]{EFEFEF}}c ccc|ccc
>{\columncolor[HTML]{EFEFEF}}c ccc|}
\hline
\cellcolor[HTML]{ECF4FF}                                        & \multicolumn{1}{c|}{\cellcolor[HTML]{ECF4FF}}                                                                    & \multicolumn{7}{c|}{\cellcolor[HTML]{ECF4FF}\textbf{Average execution Time (msec)}}                                                                                                                                                                                                                                                                                                                                                                                                                                                                                                                                                                                                                                                                                                                                                                                 & \multicolumn{7}{c|}{\cellcolor[HTML]{ECF4FF}\textbf{Line of Codes}}                                                                                                                                                                                                                                                                                                                                                                                                                                                                                                                                                                                                                                                                                                                                               \\ \cline{3-16} 
\cellcolor[HTML]{ECF4FF}                                        & \multicolumn{1}{c|}{\cellcolor[HTML]{ECF4FF}}                                                                    & \multicolumn{1}{c|}{\cellcolor[HTML]{ECF4FF}{\color[HTML]{000000} }}                                  & \multicolumn{1}{c|}{\cellcolor[HTML]{ECF4FF}{\color[HTML]{000000} }}                                                                                & \multicolumn{1}{c|}{\cellcolor[HTML]{ECF4FF}{\color[HTML]{000000} }}                                                                                & \multicolumn{1}{c|}{\cellcolor[HTML]{ECF4FF}{\color[HTML]{000000} }}                                  & \multicolumn{3}{c|}{\cellcolor[HTML]{ECF4FF}{\color[HTML]{000000} \textbf{$\Delta$}}}                                                                                                                                                                                                                                                          & \multicolumn{1}{c|}{\cellcolor[HTML]{ECF4FF}{\color[HTML]{000000} }}                                  & \multicolumn{1}{c|}{\cellcolor[HTML]{ECF4FF}{\color[HTML]{000000} }}                                                                                & \multicolumn{1}{c|}{\cellcolor[HTML]{ECF4FF}{\color[HTML]{000000} }}                                                                                & \multicolumn{1}{c|}{\cellcolor[HTML]{ECF4FF}{\color[HTML]{000000} }}                                  & \multicolumn{3}{c|}{\cellcolor[HTML]{ECF4FF}{\color[HTML]{000000} \textbf{$\Delta$}}}                                                                                                                                                                                                        \\ \cline{7-9} \cline{14-16} 
\multirow{-3}{*}{\cellcolor[HTML]{ECF4FF}\textbf{Devices}}      & \multicolumn{1}{c|}{\multirow{-3}{*}{\cellcolor[HTML]{ECF4FF}\textbf{Benchmarks}}}                                 & \multicolumn{1}{c|}{\multirow{-2}{*}{\cellcolor[HTML]{ECF4FF}{\color[HTML]{000000} \textbf{Orig}}}}   & \multicolumn{1}{c|}{\multirow{-2}{*}{\cellcolor[HTML]{ECF4FF}{\color[HTML]{000000} \textbf{\begin{tabular}[c]{@{}c@{}}PDL\\ {\cite{pendulum}}\end{tabular}}}}} & \multicolumn{1}{c|}{\multirow{-2}{*}{\cellcolor[HTML]{ECF4FF}{\color[HTML]{000000} \textbf{\begin{tabular}[c]{@{}c@{}}DFZ\\ {\cite{diffuzzar}}\end{tabular}}}}} & \multicolumn{1}{c|}{\multirow{-2}{*}{\cellcolor[HTML]{ECF4FF}{\color[HTML]{000000} \textbf{DISARM}}}} & \multicolumn{1}{c|}{\cellcolor[HTML]{ECF4FF}{\color[HTML]{000000} \textbf{\begin{tabular}[c]{@{}c@{}}Orig \\ (\%)\end{tabular}}}} & \multicolumn{1}{c|}{\cellcolor[HTML]{ECF4FF}\textbf{\begin{tabular}[c]{@{}c@{}}PDL\\  (\%)\end{tabular}}} & \cellcolor[HTML]{ECF4FF}\textbf{\begin{tabular}[c]{@{}c@{}}DFZ \\ (\%)\end{tabular}}    & \multicolumn{1}{c|}{\multirow{-2}{*}{\cellcolor[HTML]{ECF4FF}{\color[HTML]{000000} \textbf{Orig}}}}   & \multicolumn{1}{c|}{\multirow{-2}{*}{\cellcolor[HTML]{ECF4FF}{\color[HTML]{000000} \textbf{\begin{tabular}[c]{@{}c@{}}PDL\\ {\cite{pendulum}}\end{tabular}}}}} & \multicolumn{1}{c|}{\multirow{-2}{*}{\cellcolor[HTML]{ECF4FF}{\color[HTML]{000000} \textbf{\begin{tabular}[c]{@{}c@{}}DFZ\\ {\cite{diffuzzar}}\end{tabular}}}}} & \multicolumn{1}{c|}{\multirow{-2}{*}{\cellcolor[HTML]{ECF4FF}{\color[HTML]{000000} \textbf{DISARM}}}} & \multicolumn{1}{c|}{\cellcolor[HTML]{ECF4FF}{\color[HTML]{000000} \textbf{Orig}}}                       & \multicolumn{1}{c|}{\cellcolor[HTML]{ECF4FF}\textbf{PDL}}                            & \cellcolor[HTML]{ECF4FF}\textbf{DFZ}                                                 \\ \hline
{\color[HTML]{000000} }                                         & {\color[HTML]{000000} blazer\_modpow1}                                                                           & {\color[HTML]{000000} 555.89}                                                                         & {\color[HTML]{000000} 558.06}                                                                                                                       & {\color[HTML]{000000} 568.89}                                                                                                                       & {\color[HTML]{000000} 559.85}                                                                         & {\color[HTML]{000000} 0.71}                                                                                                       & 0.32                                                                                                      & -1.59                                                                                   & {\color[HTML]{000000} 30}                                                                             & {\color[HTML]{000000} 51}                                                                                                                           & {\color[HTML]{000000} 35}                                                                                                                           & {\color[HTML]{000000} 37}                                                                             & {\color[HTML]{000000} 7}                                                                                & -14                                                                                  & 2                                                                                    \\
{\color[HTML]{000000} }                                         & {\color[HTML]{000000} blazer\_array}                                                                             & {\color[HTML]{000000} 0.78}                                                                           & {\color[HTML]{000000} 2.28}                                                                                                                         & {\color[HTML]{000000} 1.48}                                                                                                                         & {\color[HTML]{000000} 1}                                                                              & {\color[HTML]{000000} 28.21}                                                                                                      & - 56.14                                                                                                   & -32.43\%                                                                                & {\color[HTML]{000000} 26}                                                                             & {\color[HTML]{000000} 45}                                                                                                                           & {\color[HTML]{000000} 34}                                                                                                                           & {\color[HTML]{000000} 47}                                                                             & {\color[HTML]{000000} 21}                                                                               & 2                                                                                    & 13                                                                                   \\
{\color[HTML]{000000} }                                         & {\color[HTML]{000000} blazer\_sanity}                                                                            & {\color[HTML]{000000} 1.13}                                                                           & {\color[HTML]{000000} 2.2}                                                                                                                          & {\color[HTML]{000000} *}                                                                                                                            & {\color[HTML]{000000} 1.13}                                                                           & {\color[HTML]{000000} 0}                                                                                                          & - 49.09                                                                                                   & $\dagger$                                                                                       & {\color[HTML]{000000} 25}                                                                             & {\color[HTML]{000000} 47}                                                                                                                           & {\color[HTML]{000000} 33}                                                                                                                           & {\color[HTML]{000000} 27}                                                                             & {\color[HTML]{000000} 2}                                                                                & -20                                                                                  & -6                                                                                   \\
{\color[HTML]{000000} }                                         & {\color[HTML]{000000} blazer\_straightline}                                                                      & {\color[HTML]{000000} 1.24}                                                                           & {\color[HTML]{000000} 2.23}                                                                                                                         & {\color[HTML]{000000} *}                                                                                                                            & {\color[HTML]{000000} 1.5}                                                                            & {\color[HTML]{000000} 28.21}                                                                                                      & -32.74                                                                                                    & $\dagger$                                                                                      & {\color[HTML]{000000} 24}                                                                             & {\color[HTML]{000000} 48}                                                                                                                           & {\color[HTML]{000000} 55}                                                                                                                           & {\color[HTML]{000000} 39}                                                                             & {\color[HTML]{000000} 15}                                                                               & -9                                                                                   & -16                                                                                  \\
{\color[HTML]{000000} }                                         & {\color[HTML]{000000} blazer\_unixlogin}                                                                         & {\color[HTML]{000000} 0.26}                                                                           & {\color[HTML]{000000} 0.35}                                                                                                                         & {\color[HTML]{000000} 0.26}                                                                                                                         & {\color[HTML]{000000} 0.29}                                                                           & {\color[HTML]{000000} 11.54}                                                                                                      & -17.14                                                                                                    & 8.61\%                                                                                  & {\color[HTML]{000000} 33}                                                                             & {\color[HTML]{000000} 37}                                                                                                                           & {\color[HTML]{000000} 59}                                                                                                                           & {\color[HTML]{000000} 66}                                                                             & {\color[HTML]{000000} 33}                                                                               & 29                                                                                   & 7                                                                                    \\
{\color[HTML]{000000} }                                         & {\color[HTML]{000000} themis\_boot}                                                                              & {\color[HTML]{000000} 1.14}                                                                           & {\color[HTML]{000000} 3.3}                                                                                                                          & {\color[HTML]{000000} 1.3}                                                                                                                          & {\color[HTML]{000000} 1.16}                                                                           & {\color[HTML]{000000} 1.75}                                                                                                       & -64.85                                                                                                    & -10.77                                                                                  & {\color[HTML]{000000} 38}                                                                             & {\color[HTML]{000000} 57}                                                                                                                           & {\color[HTML]{000000} 45}                                                                                                                           & {\color[HTML]{000000} 43}                                                                             & {\color[HTML]{000000} 5}                                                                                & -14                                                                                  & -2                                                                                   \\
{\color[HTML]{000000} }                                         & {\color[HTML]{000000} themis\_picketbox}                                                                         & {\color[HTML]{000000} 1.14}                                                                           & {\color[HTML]{000000} 1.49}                                                                                                                         & {\color[HTML]{000000} 1.16}                                                                                                                         & {\color[HTML]{000000} 1.147}                                                                          & {\color[HTML]{000000} 0.44}                                                                                                       & -23.00                                                                                                    & -1.12                                                                                   & {\color[HTML]{000000} 31}                                                                             & {\color[HTML]{000000} 44}                                                                                                                           & {\color[HTML]{000000} 46}                                                                                                                           & {\color[HTML]{000000} 42}                                                                             & {\color[HTML]{000000} 11}                                                                               & -2                                                                                   & -4                                                                                   \\
{\color[HTML]{000000} }                                         & {\color[HTML]{000000} blazer\_passwordEq}                                                                        & {\color[HTML]{000000} 1.17}                                                                           & {\color[HTML]{000000} 1.39}                                                                                                                         & {\color[HTML]{000000} 1.2}                                                                                                                          & {\color[HTML]{000000} 1.17}                                                                           & {\color[HTML]{000000} 0}                                                                                                          & -15.83                                                                                                    & -2.50                                                                                   & {\color[HTML]{000000} 33}                                                                             & {\color[HTML]{000000} 54}                                                                                                                           & {\color[HTML]{000000} 41}                                                                                                                           & {\color[HTML]{000000} 33}                                                                             & {\color[HTML]{000000} 0}                                                                                & -21                                                                                  & -8                                                                                   \\
{\color[HTML]{000000} }                                         & {\color[HTML]{000000} example\_PWCheck}                                                                          & {\color[HTML]{000000} 1.17}                                                                           & {\color[HTML]{000000} 3.28}                                                                                                                         & {\color[HTML]{000000} 1.387}                                                                                                                        & {\color[HTML]{000000} 1.2}                                                                            & {\color[HTML]{000000} 2.56}                                                                                                       & -63.41                                                                                                    & -13.48                                                                                  & {\color[HTML]{000000} 22}                                                                             & {\color[HTML]{000000} 47}                                                                                                                           & {\color[HTML]{000000} 30}                                                                                                                           & {\color[HTML]{000000} 24}                                                                             & {\color[HTML]{000000} 2}                                                                                & -23                                                                                  & -6                                                                                   \\
{\color[HTML]{000000} }                                         & {\color[HTML]{000000} themis\_jdk}                                                                               & {\color[HTML]{000000} 1.5}                                                                            & {\color[HTML]{000000} 3.38}                                                                                                                         & {\color[HTML]{000000} 1.589}                                                                                                                        & {\color[HTML]{000000} 1.53}                                                                           & {\color[HTML]{000000} 2.00}                                                                                                       & -54.73                                                                                                    & -3.71                                                                                   & {\color[HTML]{000000} 21}                                                                             & {\color[HTML]{000000} 51}                                                                                                                           & {\color[HTML]{000000} 31}                                                                                                                           & {\color[HTML]{000000} 25}                                                                             & {\color[HTML]{000000} 4}                                                                                & -26                                                                                  & -6                                                                                   \\ \cline{2-16} 
\multirow{-11}{*}{{\color[HTML]{000000} \textbf{Raspberry PI}}} & \cellcolor[HTML]{E4F5D7}\textbf{\begin{tabular}[c]{@{}l@{}}Average\\ Median\end{tabular}}                        & \cellcolor[HTML]{E4F5D7}\textbf{\begin{tabular}[c]{@{}c@{}}-\\ -\end{tabular}}                        & \cellcolor[HTML]{E4F5D7}\textbf{\begin{tabular}[c]{@{}c@{}}-\\  -\end{tabular}}                                                                     & \cellcolor[HTML]{E4F5D7}\textbf{\begin{tabular}[c]{@{}c@{}}-\\ -\end{tabular}}                                                                      & \cellcolor[HTML]{E4F5D7}\textbf{\begin{tabular}[c]{@{}c@{}}-\\ -\end{tabular}}                        & \cellcolor[HTML]{E4F5D7}\textbf{\begin{tabular}[c]{@{}c@{}}6.72\\ 1.87\end{tabular}}                                              & \cellcolor[HTML]{E4F5D7}\textbf{\begin{tabular}[c]{@{}c@{}}−37.96\\ −40.92\end{tabular}}                  & \cellcolor[HTML]{E4F5D7}\textbf{\begin{tabular}[c]{@{}c@{}}−7.12\\ −3.105\end{tabular}} & \cellcolor[HTML]{E4F5D7}\textbf{\begin{tabular}[c]{@{}c@{}}-\\ -\end{tabular}}                        & \cellcolor[HTML]{E4F5D7}\textbf{\begin{tabular}[c]{@{}c@{}}-\\ -\end{tabular}}                                                                      & \cellcolor[HTML]{E4F5D7}\textbf{\begin{tabular}[c]{@{}c@{}}-\\ -\end{tabular}}                                                                      & \cellcolor[HTML]{E4F5D7}\textbf{\begin{tabular}[c]{@{}c@{}}-\\ -\end{tabular}}                        & \cellcolor[HTML]{E4F5D7}\textbf{\begin{tabular}[c]{@{}c@{}}9.9\\ 6\end{tabular}}                        & \cellcolor[HTML]{E4F5D7}\textbf{\begin{tabular}[c]{@{}c@{}}-4.6\\ -14\end{tabular}}  & \cellcolor[HTML]{E4F5D7}\textbf{\begin{tabular}[c]{@{}c@{}}-2.6\\ -5\end{tabular}}   \\ \hline
{\color[HTML]{000000} }                                         & {\color[HTML]{000000} blazer\_modpow1}                                                                           & {\color[HTML]{000000} 421}                                                                            & {\color[HTML]{000000} 485}                                                                                                                          & {\color[HTML]{000000} 506}                                                                                                                          & {\color[HTML]{000000} 438}                                                                            & {\color[HTML]{000000} 4.04}                                                                                                       & -9.69                                                                                                     & -13.44                                                                                  & {\color[HTML]{000000} 30}                                                                             & {\color[HTML]{000000} 51}                                                                                                                           & {\color[HTML]{000000} 35}                                                                                                                           & {\color[HTML]{000000} 35}                                                                             & {\color[HTML]{000000} 5}                                                                                & -16                                                                                  & 0                                                                                    \\
{\color[HTML]{000000} }                                         & {\color[HTML]{000000} blazer\_array}                                                                             & {\color[HTML]{000000} 1.16}                                                                           & {\color[HTML]{000000} 2.8}                                                                                                                          & {\color[HTML]{000000} 1.21}                                                                                                                         & {\color[HTML]{000000} 1.18}                                                                           & {\color[HTML]{000000} 1.72}                                                                                                       & -57.85                                                                                                    & -2.47                                                                                   & {\color[HTML]{000000} 26}                                                                             & {\color[HTML]{000000} 45}                                                                                                                           & {\color[HTML]{000000} 34}                                                                                                                           & {\color[HTML]{000000} 49}                                                                             & {\color[HTML]{000000} 23}                                                                               & 4                                                                                    & 15                                                                                   \\
{\color[HTML]{000000} }                                         & {\color[HTML]{000000} blazer\_sanity}                                                                            & {\color[HTML]{000000} 1.22}                                                                           & {\color[HTML]{000000} 2.5}                                                                                                                          & {\color[HTML]{000000} *}                                                                                                                            & {\color[HTML]{000000} 1.22}                                                                           & {\color[HTML]{000000} 0}                                                                                                          & -51.2                                                                                                     & $\dagger$                                                                                      & {\color[HTML]{000000} 25}                                                                             & {\color[HTML]{000000} 47}                                                                                                                           & {\color[HTML]{000000} 33}                                                                                                                           & {\color[HTML]{000000} 26}                                                                             & {\color[HTML]{000000} 1}                                                                                & -19                                                                                  & -7                                                                                   \\
{\color[HTML]{000000} }                                         & {\color[HTML]{000000} blazer\_straightline}                                                                      & {\color[HTML]{000000} 1.24}                                                                           & {\color[HTML]{000000} 2.7}                                                                                                                          & {\color[HTML]{000000} *}                                                                                                                            & {\color[HTML]{000000} 1.57}                                                                           & {\color[HTML]{000000} 26.61}                                                                                                      & -1.85                                                                                                     & $\dagger$                                                                                      & {\color[HTML]{000000} 24}                                                                             & {\color[HTML]{000000} 48}                                                                                                                           & {\color[HTML]{000000} 55}                                                                                                                           & {\color[HTML]{000000} 38}                                                                             & {\color[HTML]{000000} 14}                                                                               & -10                                                                                  & -17                                                                                  \\
{\color[HTML]{000000} }                                         & {\color[HTML]{000000} blazer\_unixlogin}                                                                         & {\color[HTML]{000000} 0.25}                                                                           & {\color[HTML]{000000} 0.32}                                                                                                                         & {\color[HTML]{000000} 0.26}                                                                                                                         & {\color[HTML]{000000} 0.25}                                                                           & {\color[HTML]{000000} 0}                                                                                                          & -21.87                                                                                                    & -3.84                                                                                   & {\color[HTML]{000000} 33}                                                                             & {\color[HTML]{000000} 37}                                                                                                                           & {\color[HTML]{000000} 59}                                                                                                                           & {\color[HTML]{000000} 65}                                                                             & {\color[HTML]{000000} 32}                                                                               & 28                                                                                   & 6                                                                                    \\
{\color[HTML]{000000} }                                         & {\color[HTML]{000000} themis\_boot}                                                                              & {\color[HTML]{000000} 1.29}                                                                           & {\color[HTML]{000000} 3.03}                                                                                                                         & {\color[HTML]{000000} 1.36}                                                                                                                         & {\color[HTML]{000000} 1.29}                                                                           & {\color[HTML]{000000} 0}                                                                                                          & -57.42                                                                                                    & -5.14                                                                                   & {\color[HTML]{000000} 38}                                                                             & {\color[HTML]{000000} 57}                                                                                                                           & {\color[HTML]{000000} 45}                                                                                                                           & {\color[HTML]{000000} 43}                                                                             & {\color[HTML]{000000} 5}                                                                                & -14                                                                                  & -2                                                                                   \\
{\color[HTML]{000000} }                                         & {\color[HTML]{000000} themis\_picketbox}                                                                         & {\color[HTML]{000000} 1.18}                                                                           & {\color[HTML]{000000} 1.62}                                                                                                                         & {\color[HTML]{000000} 1.31}                                                                                                                         & {\color[HTML]{000000} 1.29}                                                                           & {\color[HTML]{000000} 9.32}                                                                                                       & -20.37                                                                                                    & -1.52                                                                                   & {\color[HTML]{000000} 31}                                                                             & {\color[HTML]{000000} 44}                                                                                                                           & {\color[HTML]{000000} 46}                                                                                                                           & {\color[HTML]{000000} 42}                                                                             & {\color[HTML]{000000} 11}                                                                               & -2                                                                                   & -4                                                                                   \\
{\color[HTML]{000000} }                                         & {\color[HTML]{000000} blazer\_passwordEq}                                                                        & {\color[HTML]{000000} 1.29}                                                                           & {\color[HTML]{000000} 1.53}                                                                                                                         & {\color[HTML]{000000} 1.35}                                                                                                                         & {\color[HTML]{000000} 1.29}                                                                           & {\color[HTML]{000000} 0}                                                                                                          & -15.68                                                                                                    & -4.44                                                                                   & {\color[HTML]{000000} 33}                                                                             & {\color[HTML]{000000} 54}                                                                                                                           & {\color[HTML]{000000} 41}                                                                                                                           & {\color[HTML]{000000} 33}                                                                             & {\color[HTML]{000000} 0}                                                                                & -21                                                                                  & -8                                                                                   \\
{\color[HTML]{000000} }                                         & {\color[HTML]{000000} example\_PWCheck}                                                                          & {\color[HTML]{000000} 1.2}                                                                            & {\color[HTML]{000000} 3.2}                                                                                                                          & {\color[HTML]{000000} 1.39}                                                                                                                         & {\color[HTML]{000000} 1.22}                                                                           & {\color[HTML]{000000} 1.67}                                                                                                       & -61.87                                                                                                    & -12.23                                                                                  & {\color[HTML]{000000} 22}                                                                             & {\color[HTML]{000000} 47}                                                                                                                           & {\color[HTML]{000000} 30}                                                                                                                           & {\color[HTML]{000000} 24}                                                                             & {\color[HTML]{000000} 2}                                                                                & -23                                                                                  & -6                                                                                   \\
{\color[HTML]{000000} }                                         & {\color[HTML]{000000} themis\_jdk}                                                                               & {\color[HTML]{000000} 1.55}                                                                           & {\color[HTML]{000000} 3.42}                                                                                                                         & {\color[HTML]{000000} 1.61}                                                                                                                         & {\color[HTML]{000000} 1.57}                                                                           & {\color[HTML]{000000} 1.29}                                                                                                       & -54.09                                                                                                    & -2.48                                                                                   & {\color[HTML]{000000} 21}                                                                             & {\color[HTML]{000000} 51}                                                                                                                           & {\color[HTML]{000000} 31}                                                                                                                           & {\color[HTML]{000000} 25}                                                                             & {\color[HTML]{000000} 4}                                                                                & -26                                                                                  & -6                                                                                   \\ \cline{2-16} 
\multirow{-11}{*}{{\color[HTML]{000000} \textbf{Jetson Nano}}}  & \cellcolor[HTML]{E4F5D7}\textbf{\begin{tabular}[c]{@{}l@{}}Average\\ Median\end{tabular}}                        & \cellcolor[HTML]{E4F5D7}\textbf{\begin{tabular}[c]{@{}c@{}}-\\ -\end{tabular}}                        & \cellcolor[HTML]{E4F5D7}\textbf{\begin{tabular}[c]{@{}c@{}}-\\ -\end{tabular}}                                                                      & \cellcolor[HTML]{E4F5D7}\textbf{\begin{tabular}[c]{@{}c@{}}-\\ -\end{tabular}}                                                                      & \cellcolor[HTML]{E4F5D7}\textbf{\begin{tabular}[c]{@{}c@{}}-\\ -\end{tabular}}                        & \cellcolor[HTML]{E4F5D7}\textbf{\begin{tabular}[c]{@{}c@{}}4.47\\ 1.48\end{tabular}}                                              & \cellcolor[HTML]{E4F5D7}\textbf{\begin{tabular}[c]{@{}c@{}}-39.19\\ -46.52\end{tabular}}                  & \cellcolor[HTML]{E4F5D7}\textbf{\begin{tabular}[c]{@{}c@{}}−5.70\\ −4.15\end{tabular}}  & \cellcolor[HTML]{E4F5D7}\textbf{\begin{tabular}[c]{@{}c@{}}-\\ -\end{tabular}}                        & \cellcolor[HTML]{E4F5D7}\textbf{\begin{tabular}[c]{@{}c@{}}-\\ -\end{tabular}}                                                                      & \cellcolor[HTML]{E4F5D7}\textbf{\begin{tabular}[c]{@{}c@{}}-\\ -\end{tabular}}                                                                      & \cellcolor[HTML]{E4F5D7}\textbf{\begin{tabular}[c]{@{}c@{}}-\\ -\end{tabular}}                        & \cellcolor[HTML]{E4F5D7}\textbf{\begin{tabular}[c]{@{}c@{}}9.7\\ 5\end{tabular}}                        & \cellcolor[HTML]{E4F5D7}\textbf{\begin{tabular}[c]{@{}c@{}}-9.9\\ -15\end{tabular}}  & \cellcolor[HTML]{E4F5D7}\textbf{\begin{tabular}[c]{@{}c@{}}-2.9\\ -5\end{tabular}}   \\ \hline
{\color[HTML]{000000} }                                         & {\color[HTML]{000000} blazer\_modpow1}                                                                           & {\color[HTML]{000000} 152.09}                                                                         & {\color[HTML]{000000} 162.25}                                                                                                                       & {\color[HTML]{000000} 152.45}                                                                                                                       & {\color[HTML]{000000} 152.1}                                                                          & {\color[HTML]{000000} 0.01}                                                                                                       & -6.25                                                                                                     & -0.22                                                                                   & {\color[HTML]{000000} 30}                                                                             & {\color[HTML]{000000} 51}                                                                                                                           & {\color[HTML]{000000} 35}                                                                                                                           & {\color[HTML]{000000} 35}                                                                             & {\color[HTML]{000000} 5}                                                                                & -16                                                                                  & 0                                                                                    \\
{\color[HTML]{000000} }                                         & {\color[HTML]{000000} blazer\_array}                                                                             & {\color[HTML]{000000} 0.63}                                                                           & {\color[HTML]{000000} 1.32}                                                                                                                         & {\color[HTML]{000000} 0.965}                                                                                                                        & {\color[HTML]{000000} 0.71}                                                                           & {\color[HTML]{000000} 12.70}                                                                                                      & -46.21                                                                                                    & -26.42                                                                                  & {\color[HTML]{000000} 26}                                                                             & {\color[HTML]{000000} 45}                                                                                                                           & {\color[HTML]{000000} 34}                                                                                                                           & {\color[HTML]{000000} 42}                                                                             & {\color[HTML]{000000} 16}                                                                               & -3                                                                                   & 8                                                                                    \\
{\color[HTML]{000000} }                                         & {\color[HTML]{000000} blazer\_sanity}                                                                            & {\color[HTML]{000000} 0.64}                                                                           & {\color[HTML]{000000} 1.58}                                                                                                                         & {\color[HTML]{000000} *}                                                                                                                            & {\color[HTML]{000000} 0.69}                                                                           & {\color[HTML]{000000} 7.81}                                                                                                       & -56.32                                                                                                    & $\dagger$                                                                                     & {\color[HTML]{000000} 25}                                                                             & {\color[HTML]{000000} 47}                                                                                                                           & {\color[HTML]{000000} 33}                                                                                                                           & {\color[HTML]{000000} 38}                                                                             & {\color[HTML]{000000} 13}                                                                               & 9                                                                                    & 5                                                                                    \\
{\color[HTML]{000000} }                                         & {\color[HTML]{000000} blazer\_straightline}                                                                      & {\color[HTML]{000000} 0.69}                                                                           & {\color[HTML]{000000} 1.62}                                                                                                                         & {\color[HTML]{000000} *}                                                                                                                            & {\color[HTML]{000000} 0.83}                                                                           & {\color[HTML]{000000} 20.29}                                                                                                      & -48.76                                                                                                    & $\dagger$                                                                                     & {\color[HTML]{000000} 24}                                                                             & {\color[HTML]{000000} 48}                                                                                                                           & {\color[HTML]{000000} 55}                                                                                                                           & {\color[HTML]{000000} 26}                                                                             & {\color[HTML]{000000} 2}                                                                                & -22                                                                                  & -29                                                                                  \\
{\color[HTML]{000000} }                                         & {\color[HTML]{000000} blazer\_unixlogin}                                                                         & {\color[HTML]{000000} 0.12}                                                                           & {\color[HTML]{000000} 0.14}                                                                                                                         & {\color[HTML]{000000} 0.13}                                                                                                                         & {\color[HTML]{000000} 0.12}                                                                           & {\color[HTML]{000000} 0}                                                                                                          & -25.71                                                                                                    & -20.00                                                                                  & {\color[HTML]{000000} 33}                                                                             & {\color[HTML]{000000} 37}                                                                                                                           & {\color[HTML]{000000} 59}                                                                                                                           & {\color[HTML]{000000} 64}                                                                             & {\color[HTML]{000000} 31}                                                                               & 27                                                                                   & 5                                                                                    \\
{\color[HTML]{000000} }                                         & {\color[HTML]{000000} themis\_boot}                                                                              & {\color[HTML]{000000} 0.68}                                                                           & {\color[HTML]{000000} 1.36}                                                                                                                         & {\color[HTML]{000000} 0.81}                                                                                                                         & {\color[HTML]{000000} 0.68}                                                                           & {\color[HTML]{000000} 0}                                                                                                          & -50.0                                                                                                     & -16.04                                                                                  & {\color[HTML]{000000} 38}                                                                             & {\color[HTML]{000000} 57}                                                                                                                           & {\color[HTML]{000000} 45}                                                                                                                           & {\color[HTML]{000000} 43}                                                                             & {\color[HTML]{000000} 5}                                                                                & -14                                                                                  & -2                                                                                   \\
{\color[HTML]{000000} }                                         & {\color[HTML]{000000} themis\_picketbox}                                                                         & {\color[HTML]{000000} 0.70}                                                                           & {\color[HTML]{000000} 0.82}                                                                                                                         & {\color[HTML]{000000} 0.754}                                                                                                                        & {\color[HTML]{000000} 0.73}                                                                           & {\color[HTML]{000000} 2.96}                                                                                                       & -10.97                                                                                                    & -3.18                                                                                   & {\color[HTML]{000000} 31}                                                                             & {\color[HTML]{000000} 44}                                                                                                                           & {\color[HTML]{000000} 46}                                                                                                                           & {\color[HTML]{000000} 42}                                                                             & {\color[HTML]{000000} 11}                                                                               & -2                                                                                   & -4                                                                                   \\
{\color[HTML]{000000} }                                         & {\color[HTML]{000000} blazer\_passwordEq}                                                                        & {\color[HTML]{000000} 0.71}                                                                           & {\color[HTML]{000000} 0.79}                                                                                                                         & {\color[HTML]{000000} 0.75}                                                                                                                         & {\color[HTML]{000000} 0.71}                                                                           & {\color[HTML]{000000} 0}                                                                                                          & -11.02                                                                                                    & -5.33                                                                                   & {\color[HTML]{000000} 33}                                                                             & {\color[HTML]{000000} 54}                                                                                                                           & {\color[HTML]{000000} 41}                                                                                                                           & {\color[HTML]{000000} 33}                                                                             & {\color[HTML]{000000} 0}                                                                                & 21                                                                                   & 8                                                                                    \\
{\color[HTML]{000000} }                                         & {\color[HTML]{000000} example\_PWCheck}                                                                          & {\color[HTML]{000000} 0.72}                                                                           & {\color[HTML]{000000} 1.32}                                                                                                                         & {\color[HTML]{000000} 0.76}                                                                                                                         & {\color[HTML]{000000} 0.721}                                                                          & {\color[HTML]{000000} 0.14}                                                                                                       & -45.37                                                                                                    & -5.13                                                                                   & {\color[HTML]{000000} 22}                                                                             & {\color[HTML]{000000} 47}                                                                                                                           & {\color[HTML]{000000} 30}                                                                                                                           & {\color[HTML]{000000} 24}                                                                             & {\color[HTML]{000000} 2}                                                                                & -23                                                                                  & -6                                                                                   \\
{\color[HTML]{000000} }                                         & {\color[HTML]{000000} themis\_jdk}                                                                               & {\color[HTML]{000000} 0.82}                                                                           & {\color[HTML]{000000} 1.12}                                                                                                                         & {\color[HTML]{000000} 0.91}                                                                                                                         & {\color[HTML]{000000} 0.87}                                                                           & {\color[HTML]{000000} 6.10}                                                                                                       & -22.32                                                                                                    & -4.39                                                                                   & {\color[HTML]{000000} 21}                                                                             & {\color[HTML]{000000} 51}                                                                                                                           & {\color[HTML]{000000} 31}                                                                                                                           & {\color[HTML]{000000} 25}                                                                             & {\color[HTML]{000000} 4}                                                                                & -26                                                                                  & -6                                                                                   \\ \cline{2-16} 
\multirow{-11}{*}{{\color[HTML]{000000} \textbf{JetsonAGX}}}    & \cellcolor[HTML]{E4F5D7}{\color[HTML]{000000} \textbf{\begin{tabular}[c]{@{}l@{}}Average\\ Median\end{tabular}}} & \cellcolor[HTML]{E4F5D7}{\color[HTML]{000000} \textbf{\begin{tabular}[c]{@{}c@{}}-\\ -\end{tabular}}} & \cellcolor[HTML]{E4F5D7}{\color[HTML]{000000} \textbf{\begin{tabular}[c]{@{}c@{}}-\\ -\end{tabular}}}                                               & \cellcolor[HTML]{E4F5D7}{\color[HTML]{000000} \textbf{\begin{tabular}[c]{@{}c@{}}-\\ -\end{tabular}}}                                               & \cellcolor[HTML]{E4F5D7}{\color[HTML]{000000} \textbf{\begin{tabular}[c]{@{}c@{}}-\\ -\end{tabular}}} & \cellcolor[HTML]{E4F5D7}{\color[HTML]{000000} \textbf{\begin{tabular}[c]{@{}c@{}}5\\ 1.55\end{tabular}}}                          & \cellcolor[HTML]{E4F5D7}\textbf{\begin{tabular}[c]{@{}c@{}}−32.30\\ −35.55\end{tabular}}                  & \cellcolor[HTML]{E4F5D7}\textbf{\begin{tabular}[c]{@{}c@{}}-10.09\\ -5.23\end{tabular}} & \cellcolor[HTML]{E4F5D7}{\color[HTML]{000000} \textbf{\begin{tabular}[c]{@{}c@{}}-\\ -\end{tabular}}} & \cellcolor[HTML]{E4F5D7}{\color[HTML]{000000} \textbf{\begin{tabular}[c]{@{}c@{}}-\\ -\end{tabular}}}                                               & \cellcolor[HTML]{E4F5D7}{\color[HTML]{000000} \textbf{\begin{tabular}[c]{@{}c@{}}-\\ -\end{tabular}}}                                               & \cellcolor[HTML]{E4F5D7}{\color[HTML]{000000} \textbf{\begin{tabular}[c]{@{}c@{}}-\\ -\end{tabular}}} & \cellcolor[HTML]{E4F5D7}{\color[HTML]{000000} \textbf{\begin{tabular}[c]{@{}c@{}}8.9\\ 5\end{tabular}}} & \cellcolor[HTML]{E4F5D7}\textbf{\begin{tabular}[c]{@{}c@{}}-4.9\\ -8.5\end{tabular}} & \cellcolor[HTML]{E4F5D7}\textbf{\begin{tabular}[c]{@{}c@{}}-2.1\\ -1\end{tabular}}   \\ \hline
\multicolumn{1}{|l|}{}                                          & blazer\_modpow1                                                                                                  & 65.15                                                                                                 & 78.46                                                                                                                                               & 74.23                                                                                                                                               & 65.2                                                                                                  & 0.07                                                                                                                              & -16.90                                                                                                    & -12.16                                                                                  & 30                                                                                                    & 51                                                                                                                                                  & 35                                                                                                                                                  & 35                                                                                                    & 5                                                                                                       & -16                                                                                  & 0                                                                                    \\
\multicolumn{1}{|l|}{}                                          & blazer\_array                                                                                                    & 0.13                                                                                                  & 0.44                                                                                                                                                & {\color[HTML]{000000} 0.15}                                                                                                                         & 0.15                                                                                                  & 15.38                                                                                                                             & -65.90                                                                                                    & 0                                                                                       & 26                                                                                                    & 45                                                                                                                                                  & 34                                                                                                                                                  & 32                                                                                                    & 16                                                                                                      & -3                                                                                   & 8                                                                                    \\
\multicolumn{1}{|l|}{}                                          & blazer\_sanity                                                                                                   & 0.13                                                                                                  & 0.15                                                                                                                                                & {\color[HTML]{000000} *}                                                                                                                            & 0.14                                                                                                  & 7.69                                                                                                                              & -6.66                                                                                                     & $\dagger$                                                                                  & 25                                                                                                    & 47                                                                                                                                                  & 33                                                                                                                                                  & 29                                                                                                    & 13                                                                                                      & 9                                                                                    & 5                                                                                    \\
\multicolumn{1}{|l|}{}                                          & blazer\_straightline                                                                                             & 0.15                                                                                                  & 0.3                                                                                                                                                 & {\color[HTML]{000000} *}                                                                                                                            & 0.17                                                                                                  & 13.33                                                                                                                             & -43.33                                                                                                    & $\dagger$                                                                                 & 24                                                                                                    & 48                                                                                                                                                  & 55                                                                                                                                                  & 36                                                                                                    & 12                                                                                                      & -12                                                                                  & -20                                                                                  \\
\multicolumn{1}{|l|}{}                                          & blazer\_unixlogin                                                                                                & 0.16                                                                                                  & 0.15                                                                                                                                                & 0.18                                                                                                                                                & 0.16                                                                                                  & 0                                                                                                                                 & 0                                                                                                         & -22.22                                                                                  & 33                                                                                                    & 37                                                                                                                                                  & 59                                                                                                                                                  & 56                                                                                                    & 23                                                                                                      & 19                                                                                   & -3                                                                                   \\
\multicolumn{1}{|l|}{}                                          & themis\_boot                                                                                                     & 0.13                                                                                                  & 0.37                                                                                                                                                & 0.16                                                                                                                                                & 0.14                                                                                                  & 2.94                                                                                                                              & -62.16                                                                                                    & 6.25                                                                                    & 38                                                                                                    & 57                                                                                                                                                  & 45                                                                                                                                                  & 43                                                                                                    & 5                                                                                                       & -14                                                                                  & -2                                                                                   \\
\multicolumn{1}{|l|}{}                                          & themis\_picketbox                                                                                                & 0.13                                                                                                  & 0.17                                                                                                                                                & 0.14                                                                                                                                                & 0.137                                                                                                 & 2.23                                                                                                                              & -23.03                                                                                                    & 7.14                                                                                    & 31                                                                                                    & 44                                                                                                                                                  & 46                                                                                                                                                  & 45                                                                                                    & 11                                                                                                      & -2                                                                                   & -4                                                                                   \\
\multicolumn{1}{|l|}{}                                          & blazer\_passwordEq                                                                                               & 0.14                                                                                                  & 0.17                                                                                                                                                & 0.15                                                                                                                                                & 0.14                                                                                                  & 0                                                                                                                                 & -19.99                                                                                                    & -6.66                                                                                   & 33                                                                                                    & 54                                                                                                                                                  & 41                                                                                                                                                  & 33                                                                                                    & 0                                                                                                       & 21                                                                                   & 8                                                                                    \\
\multicolumn{1}{|l|}{}                                          & example\_PWCheck                                                                                                 & 0.14                                                                                                  & 0.37                                                                                                                                                & 0.18                                                                                                                                                & 0.13                                                                                                  & 0                                                                                                                                 & -62.16                                                                                                    & -22.22                                                                                  & 22                                                                                                    & 47                                                                                                                                                  & 30                                                                                                                                                  & 24                                                                                                    & 2                                                                                                       & -23                                                                                  & -6                                                                                   \\
\multicolumn{1}{|l|}{}                                          & themis\_jdk                                                                                                      & 0.18                                                                                                  & 0.42                                                                                                                                                & 0.22                                                                                                                                                & 0.19                                                                                                  & 5.55                                                                                                                              & -54.76                                                                                                    & -13.63                                                                                  & 21                                                                                                    & 51                                                                                                                                                  & 31                                                                                                                                                  & 25                                                                                                    & 4                                                                                                       & -26                                                                                  & -6                                                                                   \\ \cline{2-16} 
\multicolumn{1}{|l|}{\multirow{-11}{*}{\textbf{Intel Core i9}}} & \cellcolor[HTML]{E4F5D7}\textbf{\begin{tabular}[c]{@{}l@{}}Average\\ Median\end{tabular}}                        & \cellcolor[HTML]{E4F5D7}\textbf{\begin{tabular}[c]{@{}c@{}}-\\ -\end{tabular}}                        & \cellcolor[HTML]{E4F5D7}\textbf{\begin{tabular}[c]{@{}c@{}}-\\ -\end{tabular}}                                                                      & \cellcolor[HTML]{E4F5D7}\textbf{\begin{tabular}[c]{@{}c@{}}-\\ -\end{tabular}}                                                                      & \cellcolor[HTML]{E4F5D7}\textbf{\begin{tabular}[c]{@{}c@{}}-\\ -\end{tabular}}                        & \cellcolor[HTML]{E4F5D7}\textbf{\begin{tabular}[c]{@{}c@{}}4.09\\ 2.58\end{tabular}}                                              & \cellcolor[HTML]{E4F5D7}\textbf{\begin{tabular}[c]{@{}c@{}}-35.49\\ -33.18\end{tabular}}                  & \cellcolor[HTML]{E4F5D7}\textbf{\begin{tabular}[c]{@{}c@{}}-7.88\\ -9.41\end{tabular}}  & \cellcolor[HTML]{E4F5D7}\textbf{\begin{tabular}[c]{@{}c@{}}-\\ -\end{tabular}}                        & \cellcolor[HTML]{E4F5D7}\textbf{\begin{tabular}[c]{@{}c@{}}-\\ -\end{tabular}}                                                                      & \cellcolor[HTML]{E4F5D7}\textbf{\begin{tabular}[c]{@{}c@{}}-\\ -\end{tabular}}                                                                      & \cellcolor[HTML]{E4F5D7}\textbf{\begin{tabular}[c]{@{}c@{}}-\\ -\end{tabular}}                        & \cellcolor[HTML]{E4F5D7}\textbf{\begin{tabular}[c]{@{}c@{}}9.1\\ 8\end{tabular}}                        & \cellcolor[HTML]{E4F5D7}\textbf{\begin{tabular}[c]{@{}c@{}}-4.7\\ -7.5\end{tabular}} & \cellcolor[HTML]{E4F5D7}\textbf{\begin{tabular}[c]{@{}c@{}}-2.0\\ -2.5\end{tabular}} \\ \hline
\bottomrule
\end{tabular}

\raggedright\footnotesize{\textit{Orig: Original program, PDL: PENDULUM \cite{pendulum}, DFZ: DifFuzzAR \cite{diffuzzar},  *: Incorrectly fixed, $\Delta$: difference between \name~ and others \\ (e.g. Average execution time of \name~ - Average execution time of PDL), $\dagger$: not applicable because the corresponding repair was incorrect or unavailable; -: raw average/median value not meaningful for that column;}}

\end{table*}

% \vspace{-0.1in}
\subsubsection{Supporting C++/C}
In the C++/C version, the process begins by parsing the main source file using ``pycparser" to build the Abstract Syntax Tree (AST). From this AST, variable names, types (including support for multidimensional arrays), and their initializers are extracted from the main() function. The specific basic block is then retrieved using a custom basic block analyzer that identifies code segments corresponding to individual basic blocks. It operates on raw C code by reading the file line by line and identifying segmentation points based on keywords such as, $\}$, $main$, $if$, $while$, and $return$. These keywords are treated as termination points for basic blocks. A list of $start\_line$ and $end\_line$ indices is used to segment the code into candidate basic blocks. Then, blocks are refined by removing redundant braces and empty lines. Building a control flow graph $\mathcal{CFG}$ (in Algo.~\ref{algo: Cycle extraction}) for a C function whose statements have already been split into basic blocks. As parsing the basic blocks contain the starting and the ending of each block, edges are added from every lexical successor.

In Algorithm~\ref{algo: Cycle extraction} for encapsulating ($EncapculateBB()$) the basic block, a separate `envelope' C template file is parsed, and variable declarations are inserted immediately after a void function, followed by a main function. The basic block itself is inserted into the for loop found in the envelope. To avoid semantic errors, the tool verifies that the block is not merely an initialization block and contains no control structures like $if$, $for$, or $while$. If the basic block accesses arrays with variable indices, loop bounds are safely modified with constraints (e.g., $i \leq 1000000~\&\&~ x \leq 1000000$). The final enveloped code is written into a new output C file. Then using $CompileBB()$ (in Algo.~\ref{algo: Cycle extraction}) function, we compile the new output C file. 
% \vspace{-0.025in}

\subsubsection{Supporting Java}
To generate the control flow graph for Java based program we need to analyze the lexical features, for example, the $keywords$, $class$, operation, etc.  In the Java version, we use the ``javalang" parser to analyze the main Java source file, extracting declared variables and their initializers, including support for arrays and multideclarator variables. Similar to the C version, a basic block is extracted from the source using a Java-specific basic block analyzer, which follows a more semantically aware approach. It reads Java files line by line similarly, but primarily focuses on braces $\{$ and $\}$ to define logical boundaries. Java’s analyzer uses a more minimal set of termination keywords and relies on an inner method split list at keywords to segment blocks when encountering control flow statements like if and while. The tool then detects which variables are referenced in the basic block by scanning tokens, carefully ignoring method names and dot-qualified symbols. To generate the $\mathcal{CFG}$ for Java, we have used the same procedure, where it iterates over all the basic blocks and adds edges according to the lexical features based on the starting and ending of the basic blocks.

In Algorithm~\ref{algo: Cycle extraction} for encapsulating ($EncapculateBB()$) the basic block for Java version, the new `envelope' Java template is parsed as plain text, and variable declarations are inserted after the $public\_static\_void$ method declaration, followed by the main function. The basic block is inserted after the for loop header. If the block is found to consist solely of variable declarations, it is skipped. Finally, the newly composed Java code is written to the output file, with proper formatting and indentation. Then, using the $CompileBB()$ function, we compile the newly created Java file to extract the CPU cycles.

\vspace{-0.11cm}
\section{Results: Java Benchmarks}
\label{java_result}
Next, we demonstrate the effectiveness of \name~ and compare it to SOTA techniques (PENDULUM, DifFuzzAR) for a set of standard Java benchmarks (see Table~\ref{tab:benchmark_deatils}). 
PENDULUM and DifFuzzAR apply device-agnostic source-level repairs; they do not use timing measurements from the target hardware; this is by their design. We have directly used the open-sourced implementations of PENDULUM and DifFuzzAR (as provided by the authors). In contrast, \name~measures timing on the actual deployment device and applies only the mitigation needed for the selected attacker model. The hardware-aware mitigation flow is a core part of \name's contribution.

We evaluate \name~on five real-world embedded devices to focus on several hardware architectures as outlined in Table~\ref{tab:my-devices}.
Based on the attacker's threat model (see Section~\ref{threat model}), we set two thresholds: timing sensitivity ($\mathcal{T}$) and minimum LoS resilience ($\mathcal{L}_{min}$), and number of traces ($\mathcal{N_T}$). Based on the threat model inspired by HASTE~\cite{chakraborty2021haste}, for `weak attack' we choose timing sensitivity ($\mathcal{T}$) = 10 for BoS severity and minimum LoS resilience ($\mathcal{L}_{min}$) = 5 for LoS severity, on the other hand for `stronger attack' we choose timing sensitivity ($\mathcal{T}$) = 5 for BoS severity and minimum LoS resilience ($\mathcal{L}_{min}$) = 10 for LoS severity. These values are used as representative experimental settings for our devices and benchmarks; however, \name~is parameterized and can support other thresholds depending on the deployment scenario.
We also consider whether an attack is able to take multiple traces for the benchmarks, so we choose number of traces ($\mathcal{N_T}$) = 10. To parse the Java code, we have used ``javalang" in all the devices. We use cloc (Count Lines of Code), a lightweight open-source CLI that scans codebases and counts code, comment, and blank lines per language.

\vspace{-0.1in}

\subsection{CPU cycles measurements}
\label{cpuCycle}
We utilize `Performance Counter', or simply $Perf$ (included in the Linux kernel under tools/perf) \cite{Perf}, to calculate $\mathcal{C}_{real}$ for all devices, as described in Algorithm~\ref{algo: Cycle extraction}. These calculated  $\mathcal{C}_{real}$ are subsequently used to assess timing vulnerabilities in the system through Algorithm~\ref{algo: HASTE-FIX} and Algorithm~\ref{algo: HASTE-FIX-Los}. Algorithm~\ref{algo: Cycle extraction} processes asset-tainted CFGs from Algorithms~\ref{algo: HASTE-FIX} and \ref{algo: HASTE-FIX-Los}, using $\mathcal{N}_{Run}=100$ for all benchmarks (see Table~\ref{tab:benchmark_deatils}).
\vspace{-0.1in}

\subsection{Utilizing DISARM to Mitigate Vulnerabilities}
We employ Algorithm~\ref{algo: HASTE-FIX} to mitigate the BoS (Branch on Secret) vulnerability, generating revised software subroutines through hardware-in-the-loop testing using benchmarks such as the unsafe modular exponentiation function (modPow1 unsafe function, see Table~\ref{tab:benchmark_deatils}), where $\mathcal{S}$ denotes the branch on the secret key bit collected from Algorithm~\ref{algo:taint_analysis}, (here we consider the worst case scenario by taking $\mathcal{S}_{taint}$, while to avoid over-fixing, \name~filters this set and repairs only variables that influence security-relevant control flow, such as branch predicates or loop bounds, and whose measured timing effect exceeds the attacker-specific threshold on the target device ($\mathcal{S}_{ctrl}$))
and $\mathcal{T}$ = 10; with $\mathcal{N}_{Run}$ = 100; $\mathcal{N_T} = 1 $. Our goal is to minimize the BoS value less than the timing sensitivity (threshold = 10). If the value is less than the threshold value, we can say that the system is secure in terms of the attack scenario as well as the underlying hardware architectures.
We use Algorithm~\ref{algo: HASTE-FIX-Los} to mitigate LoS (Loop on Secret) vulnerabilities by generating new software subroutines via hardware-in-the-loop testing, demonstrated with benchmarks (see Table~\ref{tab:benchmark_deatils}), where $\mathcal{S}$ denotes a branch on the key bit $(exponent)$. The \name~ framework was evaluated at a timing sensitivity ($\mathcal{T}$) = 50 with $\mathcal{L}_{min}$ and $\mathcal{N}_{Run}$ = 100. The goal is to maximize LoS resilience, which measures the minimum loop iterations needed to detect timing discrepancies, with higher values indicating greater security.
%\centering
\begin{figure*}[!h]
\centering

\includegraphics[width=0.95\textwidth]{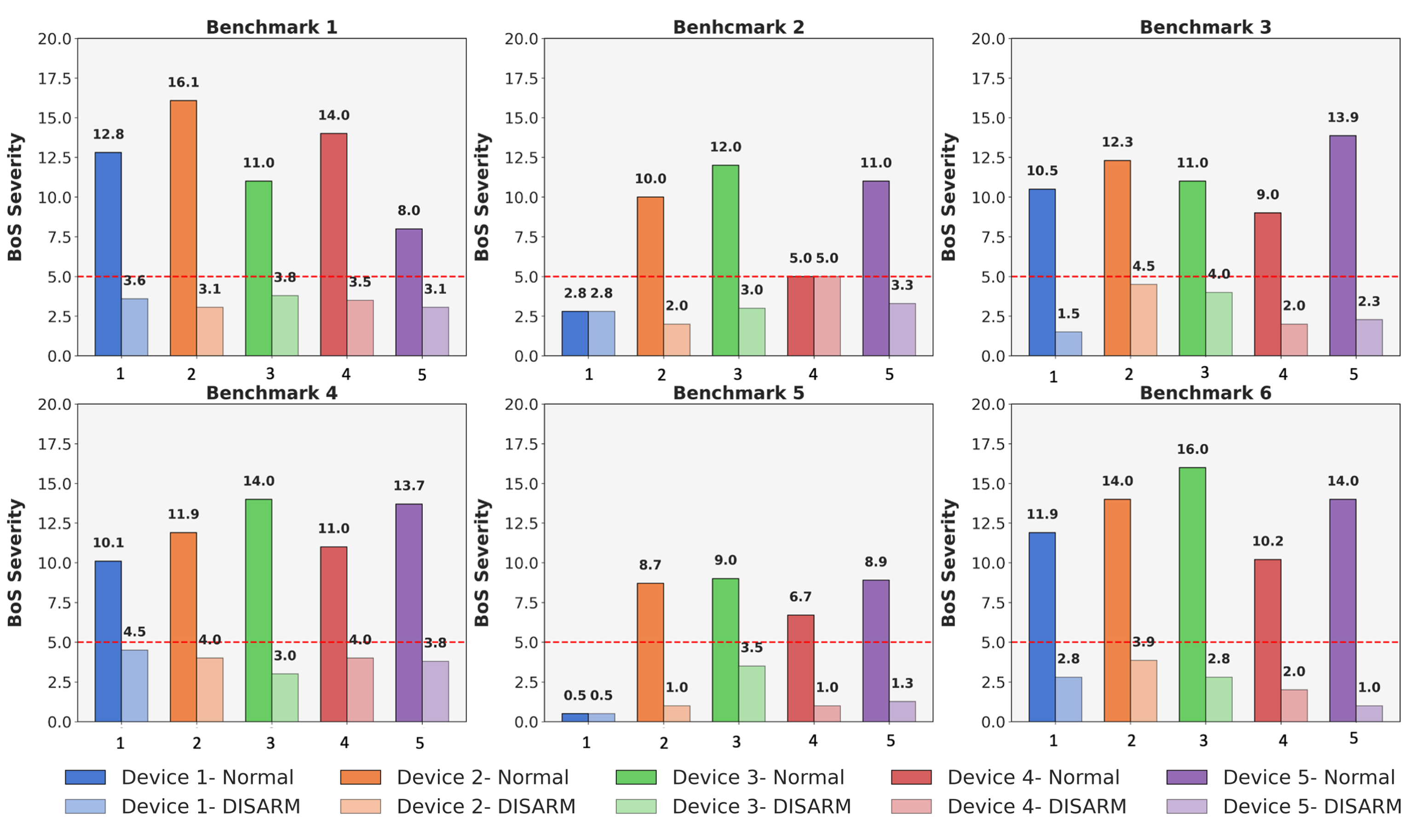}
\caption{\small{BoS Severity after fixing the code-base with \name~ (Baseline model) for timing sensitivity $\mathcal{T}$ = 5. }\label{B0S_5}}

\vspace{-0.1in}
\end{figure*}

%\centering
\begin{figure*}[!h]
\centering

\includegraphics[width=0.95\textwidth]{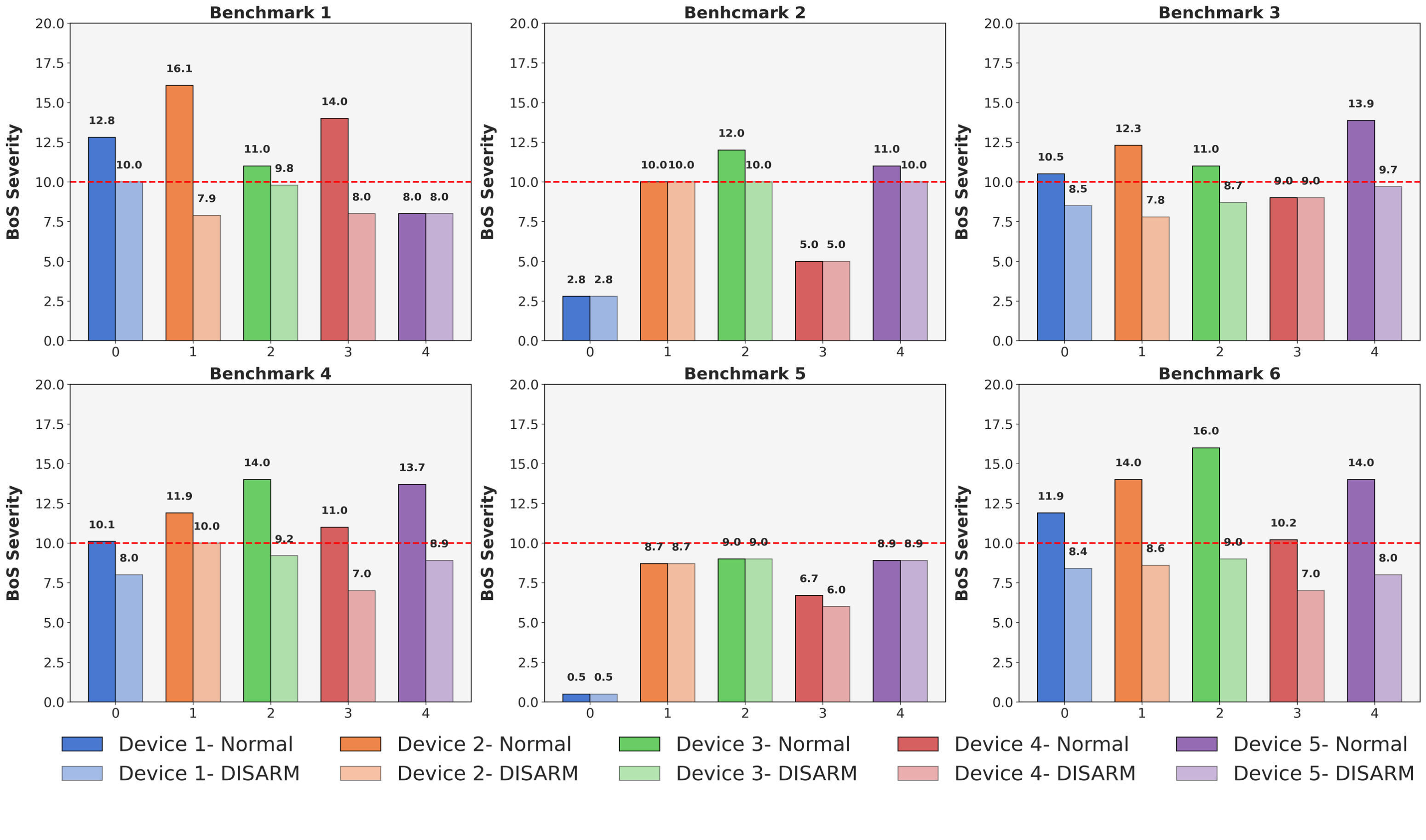}
\caption{\small{BoS Severity after fixing the code-base with \name~ (Baseline model) for timing sensitivity $\mathcal{T}$ = 10.} \label{B0S_10}}

\vspace{-0.2in}
\end{figure*}

% Required packages:
% \usepackage{multirow}
% \usepackage{graphicx}
% \usepackage{caption}
%
% Optional if \name is not already defined in your paper:
% \newcommand{\name}{DISARM}

\begin{table*}[!t]
\centering
\captionsetup{justification=centering}
\caption{BoS Severity after fixing the code-base with \name~ considering the effects of branch prediction and cache states for timing sensitivity $\mathcal{T}$ = 10; $\mathcal{B}_{t1}$ = 2 cycles; $\mathcal{B}_{t2}$ = 1 cycle and $\mathcal{P}$ = 99\%.}
\label{branching_and_Cache_effect}
\renewcommand{\arraystretch}{1.2}
\small\addtolength{\tabcolsep}{-6pt}
\newcommand{\better}[1]{\textbf{#1}}
\begin{tabular}{|
>{\columncolor[HTML]{FFFFFF}}l |
>{\columncolor[HTML]{FFFFFF}}c 
>{\columncolor[HTML]{EFEFEF}}c 
>{\columncolor[HTML]{FFFFFF}}c 
>{\columncolor[HTML]{EFEFEF}}c 
>{\columncolor[HTML]{FFFFFF}}c 
>{\columncolor[HTML]{EFEFEF}}c c
>{\columncolor[HTML]{EFEFEF}}c |
>{\columncolor[HTML]{FFFFFF}}c 
>{\columncolor[HTML]{EFEFEF}}c 
>{\columncolor[HTML]{FFFFFF}}c 
>{\columncolor[HTML]{EFEFEF}}c 
>{\columncolor[HTML]{FFFFFF}}c 
>{\columncolor[HTML]{EFEFEF}}c c
>{\columncolor[HTML]{EFEFEF}}c |}
\hline
\multicolumn{1}{|c|}{\cellcolor[HTML]{ECF4FF}} &
  \multicolumn{8}{c|}{\cellcolor[HTML]{D8F3CD}\textbf{Branch Prediction Effect}} &
  \multicolumn{8}{c|}{\cellcolor[HTML]{D8F3CD}\textbf{Cache States Effect}} \\ \cline{2-17} 
\multicolumn{1}{|c|}{\cellcolor[HTML]{ECF4FF}} &
  \multicolumn{2}{c|}{\cellcolor[HTML]{ECF4FF}\textbf{Intel Core i9}} &
  \multicolumn{2}{c|}{\cellcolor[HTML]{ECF4FF}\textbf{Jetson Nano}} &
  \multicolumn{2}{c|}{\cellcolor[HTML]{ECF4FF}\textbf{Raspberry PI}} &
  \multicolumn{2}{c|}{\cellcolor[HTML]{ECF4FF}\textbf{Jetson AGX}} &
  \multicolumn{2}{c|}{\cellcolor[HTML]{ECF4FF}\textbf{Intel Core i9}} &
  \multicolumn{2}{c|}{\cellcolor[HTML]{ECF4FF}\textbf{Jetson Nano}} &
  \multicolumn{2}{c|}{\cellcolor[HTML]{ECF4FF}\textbf{Raspberry PI}} &
  \multicolumn{2}{c|}{\cellcolor[HTML]{ECF4FF}\textbf{Jetson AGX}} \\ \cline{2-17} 
\multicolumn{1}{|c|}{\multirow{-3}{*}{\cellcolor[HTML]{ECF4FF}\textbf{Benchmark}}} &
  \multicolumn{1}{c|}{\cellcolor[HTML]{ECF4FF}\textbf{Orig}} &
  \multicolumn{1}{c|}{\cellcolor[HTML]{ECF4FF}\textbf{DISARM}} &
  \multicolumn{1}{c|}{\cellcolor[HTML]{ECF4FF}\textbf{Orig}} &
  \multicolumn{1}{c|}{\cellcolor[HTML]{ECF4FF}\textbf{DISARM}} &
  \multicolumn{1}{c|}{\cellcolor[HTML]{ECF4FF}\textbf{Orig}} &
  \multicolumn{1}{c|}{\cellcolor[HTML]{ECF4FF}\textbf{DISARM}} &
  \multicolumn{1}{c|}{\cellcolor[HTML]{ECF4FF}\textbf{Orig}} &
  \cellcolor[HTML]{ECF4FF}\textbf{DISARM} &
  \multicolumn{1}{c|}{\cellcolor[HTML]{ECF4FF}\textbf{Orig}} &
  \multicolumn{1}{c|}{\cellcolor[HTML]{ECF4FF}\textbf{DISARM}} &
  \multicolumn{1}{c|}{\cellcolor[HTML]{ECF4FF}\textbf{Orig}} &
  \multicolumn{1}{c|}{\cellcolor[HTML]{ECF4FF}\textbf{DISARM}} &
  \multicolumn{1}{c|}{\cellcolor[HTML]{ECF4FF}\textbf{Orig}} &
  \multicolumn{1}{c|}{\cellcolor[HTML]{ECF4FF}\textbf{DISARM}} &
  \multicolumn{1}{c|}{\cellcolor[HTML]{ECF4FF}\textbf{Orig}} &
  \cellcolor[HTML]{ECF4FF}\textbf{DISARM} \\ \hline
Benchmark 1 &
  13.74 & \multicolumn{1}{c|}{\cellcolor[HTML]{EFEFEF}\better{7.45}} &
  18.4 & \multicolumn{1}{c|}{\cellcolor[HTML]{EFEFEF}\better{9.34}} &
  11.35 & \multicolumn{1}{c|}{\cellcolor[HTML]{EFEFEF}\better{6.34}} &
  14.74 & \better{8.8} &
  13.67 & \multicolumn{1}{c|}{\cellcolor[HTML]{EFEFEF}\better{9.09}} &
  19 & \multicolumn{1}{c|}{\cellcolor[HTML]{EFEFEF}\better{9.56}} &
  12 & \multicolumn{1}{c|}{\cellcolor[HTML]{EFEFEF}\better{7.05}} &
  13.89 & \better{8.25} \\
Benchmark 2 &
  7.15 & \multicolumn{1}{c|}{\cellcolor[HTML]{EFEFEF}7.15} &
  12 & \multicolumn{1}{c|}{\cellcolor[HTML]{EFEFEF}\better{7.72}} &
  15.33 & \multicolumn{1}{c|}{\cellcolor[HTML]{EFEFEF}\better{9.75}} &
  8.1 & 8.1 &
  19.4 & \multicolumn{1}{c|}{\cellcolor[HTML]{EFEFEF}\better{2.3}} &
  12.35 & \multicolumn{1}{c|}{\cellcolor[HTML]{EFEFEF}\better{8.69}} &
  16.12 & \multicolumn{1}{c|}{\cellcolor[HTML]{EFEFEF}\better{9}} &
  7.69 & 7.69 \\
Benchmark 3 &
  12.33 & \multicolumn{1}{c|}{\cellcolor[HTML]{EFEFEF}\better{3.5}} &
  13.3 & \multicolumn{1}{c|}{\cellcolor[HTML]{EFEFEF}\better{8.45}} &
  17.11 & \multicolumn{1}{c|}{\cellcolor[HTML]{EFEFEF}\better{7.45}} &
  12.33 & \better{5.5} &
  14.67 & \multicolumn{1}{c|}{\cellcolor[HTML]{EFEFEF}\better{3.5}} &
  14 & \multicolumn{1}{c|}{\cellcolor[HTML]{EFEFEF}\better{9}} &
  18.98 & \multicolumn{1}{c|}{\cellcolor[HTML]{EFEFEF}\better{8.05}} &
  13.9 & \better{4.12} \\
Benchmark 4 &
  10.07 & \multicolumn{1}{c|}{\cellcolor[HTML]{EFEFEF}\better{8.15}} &
  14.9 & \multicolumn{1}{c|}{\cellcolor[HTML]{EFEFEF}\better{7.83}} &
  17.38 & \multicolumn{1}{c|}{\cellcolor[HTML]{EFEFEF}\better{6.5}} &
  12.50 & \better{8.15} &
  11.83 & \multicolumn{1}{c|}{\cellcolor[HTML]{EFEFEF}\better{9.66}} &
  15.12 & \multicolumn{1}{c|}{\cellcolor[HTML]{EFEFEF}\better{9.14}} &
  18 & \multicolumn{1}{c|}{\cellcolor[HTML]{EFEFEF}\better{6.8}} &
  10.07 & \better{8.15} \\
Benchmark 5 &
  3.37 & \multicolumn{1}{c|}{\cellcolor[HTML]{EFEFEF}3.37} &
  10.7 & \multicolumn{1}{c|}{\cellcolor[HTML]{EFEFEF}\better{6.5}} &
  10.98 & \multicolumn{1}{c|}{\cellcolor[HTML]{EFEFEF}\better{9.13}} &
  7.8 & 7.8 &
  6.87 & \multicolumn{1}{c|}{\cellcolor[HTML]{EFEFEF}6.87} &
  11.3 & \multicolumn{1}{c|}{\cellcolor[HTML]{EFEFEF}\better{7.45}} &
  12.25 & \multicolumn{1}{c|}{\cellcolor[HTML]{EFEFEF}\better{9.5}} &
  3.37 & 3.37 \\
Benchmark 6 &
  10 & \multicolumn{1}{c|}{\cellcolor[HTML]{EFEFEF}10} &
  16.8 & \multicolumn{1}{c|}{\cellcolor[HTML]{EFEFEF}\better{9.5}} &
  18 & \multicolumn{1}{c|}{\cellcolor[HTML]{EFEFEF}\better{10}} &
  13.14 & \better{9.81} &
  11.9 & \multicolumn{1}{c|}{\cellcolor[HTML]{EFEFEF}\better{8.7}} &
  17.25 & \multicolumn{1}{c|}{\cellcolor[HTML]{EFEFEF}\better{9.85}} &
  18.54 & \multicolumn{1}{c|}{\cellcolor[HTML]{EFEFEF}\better{9.3}} &
  10 & 10 \\
Benchmark 7 &
  13.69 & \multicolumn{1}{c|}{\cellcolor[HTML]{EFEFEF}\better{5.86}} &
  18.75 & \multicolumn{1}{c|}{\cellcolor[HTML]{EFEFEF}\better{9.6}} &
  26.5 & \multicolumn{1}{c|}{\cellcolor[HTML]{EFEFEF}\better{8.56}} &
  15.28 & \better{9.2} &
  13.17 & \multicolumn{1}{c|}{\cellcolor[HTML]{EFEFEF}\better{8.99}} &
  19.18 & \multicolumn{1}{c|}{\cellcolor[HTML]{EFEFEF}\better{9.78}} &
  27 & \multicolumn{1}{c|}{\cellcolor[HTML]{EFEFEF}\better{9.12}} &
  13.69 & \better{5.86} \\
Benchmark 8 &
  12.36 & \multicolumn{1}{c|}{\cellcolor[HTML]{EFEFEF}\better{7.26}} &
  24.24 & \multicolumn{1}{c|}{\cellcolor[HTML]{EFEFEF}\better{7.8}} &
  21.88 & \multicolumn{1}{c|}{\cellcolor[HTML]{EFEFEF}\better{7.88}} &
  17.28 & \better{8.19} &
  13.56 & \multicolumn{1}{c|}{\cellcolor[HTML]{EFEFEF}\better{8.96}} &
  24.98 & \multicolumn{1}{c|}{\cellcolor[HTML]{EFEFEF}\better{8.25}} &
  22.54 & \multicolumn{1}{c|}{\cellcolor[HTML]{EFEFEF}\better{8.56}} &
  12.36 & \better{7.26} \\
Benchmark 9 &
  3.05 & \multicolumn{1}{c|}{\cellcolor[HTML]{EFEFEF}3.05} &
  5.86 & \multicolumn{1}{c|}{\cellcolor[HTML]{EFEFEF}5.86} &
  11 & \multicolumn{1}{c|}{\cellcolor[HTML]{EFEFEF}\better{6.8}} &
  5.10 & 5.10 &
  2.61 & \multicolumn{1}{c|}{\cellcolor[HTML]{EFEFEF}2.61} &
  6.5 & \multicolumn{1}{c|}{\cellcolor[HTML]{EFEFEF}6.5} &
  12.56 & \multicolumn{1}{c|}{\cellcolor[HTML]{EFEFEF}\better{7.8}} &
  3.05 & 3.05 \\
Benchmark 10 &
  3.16 & \multicolumn{1}{c|}{\cellcolor[HTML]{EFEFEF}3.16} &
  4.45 & \multicolumn{1}{c|}{\cellcolor[HTML]{EFEFEF}4.45} &
  9.83 & \multicolumn{1}{c|}{\cellcolor[HTML]{EFEFEF}9.83} &
  4.00 & 4.00 &
  4.73 & \multicolumn{1}{c|}{\cellcolor[HTML]{EFEFEF}4.73} &
  4.65 & \multicolumn{1}{c|}{\cellcolor[HTML]{EFEFEF}4.65} &
  10.12 & \multicolumn{1}{c|}{\cellcolor[HTML]{EFEFEF}\better{9.5}} &
  3.16 & 3.16 \\
Benchmark 11 &
  2.06 & \multicolumn{1}{c|}{\cellcolor[HTML]{EFEFEF}2.06} &
  2.86 & \multicolumn{1}{c|}{\cellcolor[HTML]{EFEFEF}2.86} &
  5.02 & \multicolumn{1}{c|}{\cellcolor[HTML]{EFEFEF}5.02} &
  3.35 & 3.35 &
  2.16 & \multicolumn{1}{c|}{\cellcolor[HTML]{EFEFEF}2.16} &
  3.56 & \multicolumn{1}{c|}{\cellcolor[HTML]{EFEFEF}3.56} &
  6.89 & \multicolumn{1}{c|}{\cellcolor[HTML]{EFEFEF}6.89} &
  2.06 & 2.06 \\
Benchmark 12 &
  3.96 & \multicolumn{1}{c|}{\cellcolor[HTML]{EFEFEF}3.96} &
  3.33 & \multicolumn{1}{c|}{\cellcolor[HTML]{EFEFEF}3.33} &
  3.33 & \multicolumn{1}{c|}{\cellcolor[HTML]{EFEFEF}3.33} &
  3.25 & 3.25 &
  2.89 & \multicolumn{1}{c|}{\cellcolor[HTML]{EFEFEF}2.89} &
  4.05 & \multicolumn{1}{c|}{\cellcolor[HTML]{EFEFEF}4.05} &
  3.56 & \multicolumn{1}{c|}{\cellcolor[HTML]{EFEFEF}3.56} &
  3.96 & 3.96 \\ \hline
\end{tabular}
\end{table*}

\vspace{-0.15 in}
\subsection{Comparing DISARM with PENDULUM and DIfFuzzAR}
PENDULUM and DifFuzzAR (SOTA) repair timing leaks mainly by applying source code transformations that are independent of the target hardware. Their goal is to make secret-dependent code paths more balanced or to convert the program into a safer single-exit structure. In contrast, \name~ first measures the execution time on the actual target device and then uses the selected attacker model to decide how much padding is really needed. Therefore, \name~adds padding only until the measured BoS or LoS leakage falls below the security threshold for that specific hardware. As a result, \name’s lower runtime and code size overhead do not mean that all tools use the same repair strategy. Instead, they show that hardware calibrated and threat-model aware repair can avoid some unnecessary extra work introduced by fully device agnostic source level balancing.
After performing algorithms (Algorithm ~\ref{algo: HASTE-FIX} \& Algorithm ~\ref{algo: HASTE-FIX-Los}), we compared our automated repaired programs with SOTA techniques PENDULUM and DifFuzzAR in Table~\ref{tab:comparisonJava} to evaluate the time and space impact on different devices. To make the correctness comparison explicit, each repaired program is checked using three criteria: successful compilation, preservation of the original program behavior under regression testing, and satisfaction of the configured BoS/LoS security threshold. In Table~\ref{tab:comparisonJava}, ``*'' indicates an incorrect fix, i.e., the repaired program did not preserve the original behavior under regression testing. In contrast, all \name-fixed programs were successfully compiled, passed regression tests and met the configured BoS/LoS threshold, supporting the 100\% \% correctness claim for \name.
% Here we choose four different hardware architectures: (1) Raspberry PI: Quad-core Cortex-A72, (2) Jetson Nano ARM Cortex-A57;  (3) AGX Xavier Carmel ARM v8.2, and (4) Intel Core i9. 
We chose 10 benchmarks for Java, which are both common in PENDULUM and DifFuzzAR, for a better perception of the extensive analysis. In this table, the average execution time column compares the runtime of the regression testsuit between \name-fixed code and the original program, the PENDULUM-fixed program and the DifFuzzAR-fixed program. $\Delta$Orig column shows the percentage difference with the original program and \name-fixed program, which is 6.72\% on average and 1.87\% on median for Raspberry PI, which indicates that the program fixed by \name~is only 6.72\% slower than the original code. For Jetson nano $\Delta$Orig is 4.47\% on average with 1.48\% on median. Column $\Delta$PDL reports the percentage execution-time difference between \name{}-fixed and PENDULUM-fixed code; negative values indicate lower execution time for \name. Its average ranges from -32\% to -39\% across devices, showing that \name~is nearly 40\% faster than PENDULUM. Similarly, $\Delta$DFZ compares \name{} with DifFuzzAR-fixed code, averaging -5.70\% (Jetson Nano) to -10.09\% (Jetson AGX). For space overhead, $\Delta$Orig reports the line-of-code difference between the \name{}-fixed and original code.
% The average difference of the lines is 9.9, 9.7, 8.9, 9.1 for Raspberry PI, Jetson Nano, Jetson AGX and Intel Core i9, respectively. That explains the fixed code in almost 9.9, 9.7, 8.9 and 9.1 lines larger than the original code for Raspberry PI, Jetson Nano, Jetson AGX and Intel Core i9, in that order.  
\begin{takeaway}
\textbf{Takeaway}: \name~mitigates BoS/LoS with low overhead vs.\ original (median $\leq$ 2\%), while running $\approx$ 40\% faster than PENDULUM and $\approx$ 6--10\% faster than DifFuzzAR, and using fewer lines than both across devices.
\end{takeaway}
% \vspace{-0.2in}
The column $\Delta$PDL is the difference of lines with the PENDULUM-fixed code, which is -9.9 in average and -15 in median for Jetson Nano, shows that the fixed code by \name~ is taking 9.9 fewer lines than PENDULUM-fixed code. For other devices the average values are almost -5. The $\Delta$DFZ represents the variations of lines with DifFuzzAR-fixed code. 
% For each devices the average line difference is -2.0 to -2.9 which means that code fixed by \name~ has almost 2 to 2.9 line less than the DifFuzzAR-fixed code. 
Based on the performance analysis (Time and space), we can say that our framework outperforms the SOTA techniques. 

%\centering
\begin{figure*}[t]
\centering

\includegraphics[width=0.90\textwidth]{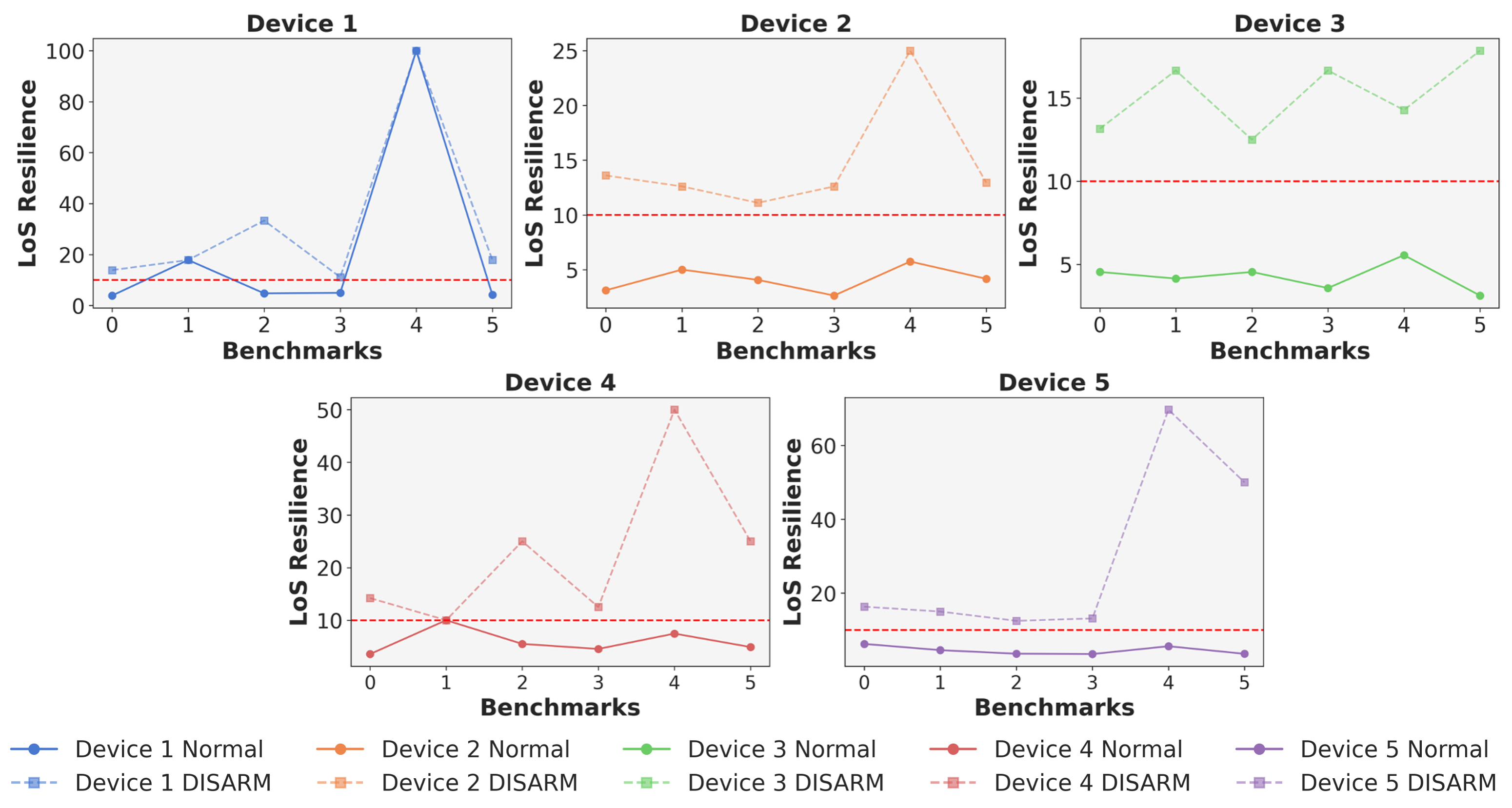}
\caption{\small{LoS Severity after \name~ for timing sensitivity 50 and minimum LoS Resilience is 10 with variable timing effect. \label{LOS_10}}}

\vspace{-0.1in}
\end{figure*}

%\centering
\begin{figure*}[ht]
\centering

\includegraphics[width=0.93\textwidth]{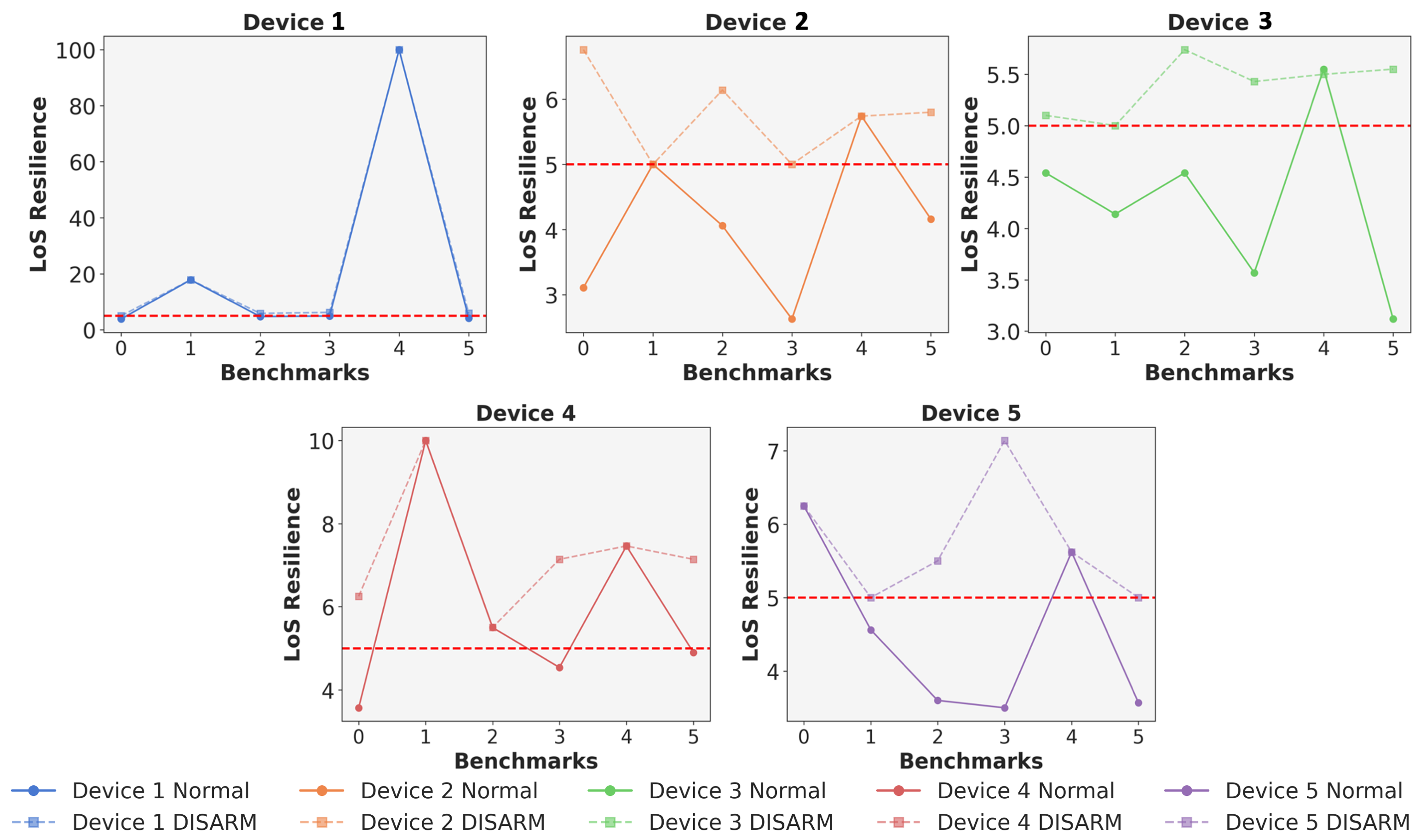}
\captionsetup{justification=centering}
\caption{\small{LoS Severity after \name~ for timing sensitivity 50 and minimum LoS Resilience is 5 with variable timing effect.}  \label{LOS_5}}

\vspace{-0.1in}
\end{figure*}

% \subsection{\textcolor{blue}{Fairness of Comparison}}
% \textcolor{blue}{% For the Java benchmarks, we use the repaired versions released by PENDULUM and DifFuzzAR where available and evaluate all programs under the same devices, inputs, and measurement protocol. 
% PENDULUM and DifFuzzAR apply device-agnostic source-level repairs; they do not use timing measurements from the target hardware. We have directly used the open-sourced implementations of PENDULUM and DifFuzzAR (as provided by the authors).
% % This is important because timing leakage and repair overhead can vary across hardware architectures due to differences in branch behavior, cache effects, and execution latency. 
% In contrast, DISARM measures timing on the actual deployment device and applies only the mitigation needed for the selected attacker model. The hardware-aware mitigation flow is a core part of DISARM's contribution. 
% % For C/C++ benchmarks, since PENDULUM and DifFuzzAR do not provide native C/C++ repair tools, we use faithful implementations of their published repair strategies only as a secondary comparison. Thus, the primary SOTA comparison is the Java evaluation in Table~\ref{tab:comparisonJava}
% }
\section {Results: C++/C Benchmarks}
We evaluate \name~ on five real-world embedded devices to focus on several hardware architectures as outlined in Table~\ref{tab:my-devices} and utilizing the same CPU-cycle measurement techniques as specified for the Java benchmarks (see Section~\ref{cpuCycle}). 
Based on the attacker's threat model (see Section~\ref{threat model}), we set two thresholds: timing sensitivity ($\mathcal{T}$) and minimum LoS resilience ($\mathcal{L}_{min}$), and number of traces ($\mathcal{N_T}$). Based on the threat model inspired by HASTE~\cite{chakraborty2021haste}, for `weak attack' we choose timing sensitivity ($\mathcal{T}$) = 10 for BoS severity and minimum LoS resilience ($\mathcal{L}_{min}$) = 5 for LoS severity, on the other hand for `stronger attack' we choose timing sensitivity ($\mathcal{T}$) = 5 for BoS severity and minimum LoS resilience ($\mathcal{L}_{min}$) = 10 for LoS severity. These values are used as representative experimental settings for our devices and benchmarks; however, \name~is parameterized and can support other thresholds depending on the deployment scenario.

PENDULUM \cite{pendulum} and DifFuzzAR \cite{diffuzzar} native-implementations do not directly support C++/C source code. Hence, for C/C++ benchmarks, we have created a nearly faithful implementation of their published repair strategies only as a secondary comparison (primary comparison was done on Java Benchmarks in Section~\ref{java_result}, earlier).

% Hence, we perform a self-comparison between different \name~ settings.  
% we built a similar constant-time mitigation tool for C/C++ and compared it with \name~across different benchmarks.

% \subsubsection{Capturing the Effects of Branch Prediction}
% To consider the impact of branch prediction, for each branch decision we consider a $\mathcal{P}$ = 99\% probability (an user-input that can be empirically determined) that the right branch will be choose. This is simulated in our framework by splitting each branch line into two pathways: one with minimal delay $\mathcal{B}_{t2}$ (right prediction); the other one with a longer delay $\mathcal{B}_{t1}$ (wrong prediction). We choose  $\mathcal{B}_{t1}$ = 2 cycles and $\mathcal{B}_{t2}$ = 1 cycle for the reported experimental results. These values can be user-determined in our automation framework.

\subsubsection{Capturing the Effects of Branch Prediction}
To consider the impact of branch prediction, for each branch decision we consider a probability $\mathcal{P}$ (a user-input that can be empirically determined) that the right branch will be chosen. This is simulated in our framework by splitting each branch line into two pathways: one with minimal delay $\mathcal{B}_{t2}$ (right prediction); the other one with a longer delay $\mathcal{B}_{t1}$ (wrong prediction). We choose $\mathcal{P} = 99\%$, $\mathcal{B}_{t1}$ = 2 cycles, and $\mathcal{B}_{t2}$ = 1 cycle for the reported experimental results. These values can be user-determined in our automation framework.

% To capture branching effects when minimizing BoS severity as well the LoS severity, we assign baseline execution times of $\mathcal{B}_{t1}$ = 2 cycles and $\mathcal{B}_{t2}$ = 1 cycle to the two successors of each infused two-way branch, with the longer path chosen with probability $\mathcal{P}$ = 99\%. We use this as an user input, but the user could profile the targeted hardware device and capture the branch time. 
% \vspace{-1 in}
\subsubsection{Capturing the Effects of Different Cache States}
To consider the effect of different cache states across different embedded devices, each instrumented basic block randomly samples a subset of its local variables and 'warms' them by accessing the respective cache line immediately before the timing window; the warm set is resampled on every run from a fresh seed. This step is carried out for each run to calculate the average basic block runtime ($\mathcal{N}_{Run}=100$, in Algorithm~\ref{algo: Cycle extraction}). 

% We further evaluated the efficacy of \name~in terms of fixing C++/C source codes with runtime vulnerabilities. 

\vspace{-0.1in}

% \input{fig_latex/multiple}

% \subsubsection{CPU cycles measurements}
% We utilize `Performance Counter', or simply $Perf$ (included in the Linux kernel under tools/perf) \cite{Perf}, to calculate $\mathcal{C}_{real}$ for all device, as described in Algorithm~\ref{algo: Cycle extraction} similarly as described earlier. 

% These calculated  $\mathcal{C}_{real}$ are subsequently used to assess timing vulnerabilities in the system through Algorithm~\ref{algo: HASTE-FIX} and Algorithm~\ref{algo: HASTE-FIX-Los}. Algorithm~\ref{algo: Cycle extraction} processes asset-tainted CFGs from Algorithms~\ref{algo: HASTE-FIX} and \ref{algo: HASTE-FIX-Los}, using $\mathcal{N}_{Run}=100$.
% \vspace{-0.1in}
% \usepackage{multirow}
% \usepackage{graphicx}
\begin{table*}[t]
\centering
\captionsetup{justification=centering}
\caption{LoS Severity after \name~ fixes the code-base accounting for the effects of branch prediction and cache states. where, timing sensitivity $\mathcal{T}$ = 50, minimum LoS Resilience $\mathcal{L}_{min}$ = 5 ; $\mathcal{B}_{t1}$ = 2 cycles ; $\mathcal{B}_{t2}$ = 1 cycle and $\mathcal{P}$ = 99\%.}
\label{branching_and_Cache_effect_LOS}
\renewcommand{\arraystretch}{1.2}
\small\addtolength{\tabcolsep}{-6pt}
\newcommand{\better}[1]{\textbf{#1}}
\begin{tabular}{|
>{\columncolor[HTML]{FFFFFF}}l |
>{\columncolor[HTML]{FFFFFF}}c 
>{\columncolor[HTML]{EFEFEF}}c 
>{\columncolor[HTML]{FFFFFF}}c 
>{\columncolor[HTML]{EFEFEF}}c 
>{\columncolor[HTML]{FFFFFF}}c 
>{\columncolor[HTML]{EFEFEF}}c c
>{\columncolor[HTML]{EFEFEF}}c |
>{\columncolor[HTML]{FFFFFF}}c 
>{\columncolor[HTML]{EFEFEF}}c 
>{\columncolor[HTML]{FFFFFF}}c 
>{\columncolor[HTML]{EFEFEF}}c 
>{\columncolor[HTML]{FFFFFF}}c 
>{\columncolor[HTML]{EFEFEF}}c c
>{\columncolor[HTML]{EFEFEF}}c |}
\hline
\multicolumn{1}{|c|}{\cellcolor[HTML]{ECF4FF}} &
  \multicolumn{8}{c|}{\cellcolor[HTML]{D8F3CD}\textbf{Branch Prediction Effect}} &
  \multicolumn{8}{c|}{\cellcolor[HTML]{D8F3CD}\textbf{Cache States Effect}} \\ \cline{2-17} 
\multicolumn{1}{|c|}{\cellcolor[HTML]{ECF4FF}} &
  \multicolumn{2}{c|}{\cellcolor[HTML]{ECF4FF}\textbf{Intel Core i9}} &
  \multicolumn{2}{c|}{\cellcolor[HTML]{ECF4FF}\textbf{Jetson Nano}} &
  \multicolumn{2}{c|}{\cellcolor[HTML]{ECF4FF}\textbf{Raspberry PI}} &
  \multicolumn{2}{c|}{\cellcolor[HTML]{ECF4FF}\textbf{Jetson AGX}} &
  \multicolumn{2}{c|}{\cellcolor[HTML]{ECF4FF}\textbf{Intel Core i9}} &
  \multicolumn{2}{c|}{\cellcolor[HTML]{ECF4FF}\textbf{Jetson Nano}} &
  \multicolumn{2}{c|}{\cellcolor[HTML]{ECF4FF}\textbf{Raspberry PI}} &
  \multicolumn{2}{c|}{\cellcolor[HTML]{ECF4FF}\textbf{Jetson AGX}} \\ \cline{2-17} 
\multicolumn{1}{|c|}{\multirow{-3}{*}{\cellcolor[HTML]{ECF4FF}\textbf{Benchmarks}}} &
  \multicolumn{1}{c|}{\cellcolor[HTML]{ECF4FF}\textbf{Orig}} &
  \multicolumn{1}{c|}{\cellcolor[HTML]{ECF4FF}\textbf{DISARM}} &
  \multicolumn{1}{c|}{\cellcolor[HTML]{ECF4FF}\textbf{Orig}} &
  \multicolumn{1}{c|}{\cellcolor[HTML]{ECF4FF}\textbf{DISARM}} &
  \multicolumn{1}{c|}{\cellcolor[HTML]{ECF4FF}\textbf{Orig}} &
  \multicolumn{1}{c|}{\cellcolor[HTML]{ECF4FF}\textbf{DISARM}} &
  \multicolumn{1}{c|}{\cellcolor[HTML]{ECF4FF}\textbf{Orig}} &
  \cellcolor[HTML]{ECF4FF}\textbf{DISARM} &
  \multicolumn{1}{c|}{\cellcolor[HTML]{ECF4FF}\textbf{Orig}} &
  \multicolumn{1}{c|}{\cellcolor[HTML]{ECF4FF}\textbf{DISARM}} &
  \multicolumn{1}{c|}{\cellcolor[HTML]{ECF4FF}\textbf{Orig}} &
  \multicolumn{1}{c|}{\cellcolor[HTML]{ECF4FF}\textbf{DISARM}} &
  \multicolumn{1}{c|}{\cellcolor[HTML]{ECF4FF}\textbf{Orig}} &
  \multicolumn{1}{c|}{\cellcolor[HTML]{ECF4FF}\textbf{DISARM}} &
  \multicolumn{1}{c|}{\cellcolor[HTML]{ECF4FF}\textbf{Orig}} &
  \cellcolor[HTML]{ECF4FF}\textbf{DISARM} \\ \hline
Benchmark 1 &
  3.63 &
  \multicolumn{1}{c|}{\cellcolor[HTML]{EFEFEF}\better{6.71}} &
  2.71 &
  \multicolumn{1}{c|}{\cellcolor[HTML]{EFEFEF}\better{5.34}} &
  7.89 &
  \multicolumn{1}{c|}{\cellcolor[HTML]{EFEFEF}\better{6.34}} &
  3.63 &
  \better{6.71} &
  3.65 &
  \multicolumn{1}{c|}{\cellcolor[HTML]{EFEFEF}\better{5.5}} &
  2.63 &
  \multicolumn{1}{c|}{\cellcolor[HTML]{EFEFEF}\better{5.23}} &
  4.16 &
  \multicolumn{1}{c|}{\cellcolor[HTML]{EFEFEF}\better{7.09}} &
  3.59 &
  \better{6.06} \\
Benchmark 2 &
  6.99 &
  \multicolumn{1}{c|}{\cellcolor[HTML]{EFEFEF}6.99} &
  4.16 &
  \multicolumn{1}{c|}{\cellcolor[HTML]{EFEFEF}\better{6.47}} &
  5.1 &
  \multicolumn{1}{c|}{\cellcolor[HTML]{EFEFEF}\better{9.75}} &
  6.99 &
  6.99 &
  2.57 &
  \multicolumn{1}{c|}{\cellcolor[HTML]{EFEFEF}\better{21.73}} &
  4.04 &
  \multicolumn{1}{c|}{\cellcolor[HTML]{EFEFEF}\better{5.75}} &
  3.1 &
  \multicolumn{1}{c|}{\cellcolor[HTML]{EFEFEF}\better{5.5}} &
  6.5 &
  6.5 \\
Benchmark 3 &
  4.05 &
  \multicolumn{1}{c|}{\cellcolor[HTML]{EFEFEF}\better{14.28}} &
  3.75 &
  \multicolumn{1}{c|}{\cellcolor[HTML]{EFEFEF}\better{5.91}} &
  6.71 &
  \multicolumn{1}{c|}{\cellcolor[HTML]{EFEFEF}\better{7.45}} &
  4.16 &
  \better{14.28} &
  3.4 &
  \multicolumn{1}{c|}{\cellcolor[HTML]{EFEFEF}\better{14.28}} &
  3.57 &
  \multicolumn{1}{c|}{\cellcolor[HTML]{EFEFEF}\better{5.5}} &
  2.63 &
  \multicolumn{1}{c|}{\cellcolor[HTML]{EFEFEF}\better{6.21}} &
  3.59 &
  \better{12.13} \\
Benchmark 4 &
  4.96 &
  \multicolumn{1}{c|}{\cellcolor[HTML]{EFEFEF}\better{8.15}} &
  3.35 &
  \multicolumn{1}{c|}{\cellcolor[HTML]{EFEFEF}\better{6.38}} &
  6.5 &
  \multicolumn{1}{c|}{\cellcolor[HTML]{EFEFEF}6.5} &
  4.96 &
  \better{6.14} &
  4.22 &
  \multicolumn{1}{c|}{\cellcolor[HTML]{EFEFEF}\better{5.17}} &
  3.3 &
  \multicolumn{1}{c|}{\cellcolor[HTML]{EFEFEF}\better{5.47}} &
  2.77 &
  \multicolumn{1}{c|}{\cellcolor[HTML]{EFEFEF}\better{7.35}} &
  4.96 &
  \better{6.13} \\
Benchmark 5 &
  14.83 &
  \multicolumn{1}{c|}{\cellcolor[HTML]{EFEFEF}14.83} &
  4.67 &
  \multicolumn{1}{c|}{\cellcolor[HTML]{EFEFEF}\better{7.69}} &
  5.47 &
  \multicolumn{1}{c|}{\cellcolor[HTML]{EFEFEF}\better{9.13}} &
  14.83 &
  14.83 &
  7.27 &
  \multicolumn{1}{c|}{\cellcolor[HTML]{EFEFEF}7.27} &
  4.42 &
  \multicolumn{1}{c|}{\cellcolor[HTML]{EFEFEF}\better{6.71}} &
  4.08 &
  \multicolumn{1}{c|}{\cellcolor[HTML]{EFEFEF}\better{5.25}} &
  14.83 &
  14.83 \\
Benchmark 6 &
  5 &
  \multicolumn{1}{c|}{\cellcolor[HTML]{EFEFEF}5} &
  2.97 &
  \multicolumn{1}{c|}{\cellcolor[HTML]{EFEFEF}\better{5.26}} &
  4.5 &
  \multicolumn{1}{c|}{\cellcolor[HTML]{EFEFEF}\better{10}} &
  5 &
  5 &
  4.2 &
  \multicolumn{1}{c|}{\cellcolor[HTML]{EFEFEF}\better{5.74}} &
  2.89 &
  \multicolumn{1}{c|}{\cellcolor[HTML]{EFEFEF}\better{5.05}} &
  2.69 &
  \multicolumn{1}{c|}{\cellcolor[HTML]{EFEFEF}\better{5.37}} &
  5 &
  5 \\
Benchmark 7 &
  3.65 &
  \multicolumn{1}{c|}{\cellcolor[HTML]{EFEFEF}\better{8.53}} &
  2.6 &
  \multicolumn{1}{c|}{\cellcolor[HTML]{EFEFEF}\better{5.2}} &
  5.84 &
  \multicolumn{1}{c|}{\cellcolor[HTML]{EFEFEF}\better{8.56}} &
  3.65 &
  \better{8.53} &
  3.79 &
  \multicolumn{1}{c|}{\cellcolor[HTML]{EFEFEF}\better{5.59}} &
  2.6 &
  \multicolumn{1}{c|}{\cellcolor[HTML]{EFEFEF}\better{5.11}} &
  1.85 &
  \multicolumn{1}{c|}{\cellcolor[HTML]{EFEFEF}\better{5.48}} &
  3.58 &
  \better{8.53} \\
Benchmark 8 &
  4.04 &
  \multicolumn{1}{c|}{\cellcolor[HTML]{EFEFEF}\better{6.88}} &
  2.06 &
  \multicolumn{1}{c|}{\cellcolor[HTML]{EFEFEF}\better{6.41}} &
  6.34 &
  \multicolumn{1}{c|}{\cellcolor[HTML]{EFEFEF}\better{7.88}} &
  4.04 &
  \better{6.88} &
  3.68 &
  \multicolumn{1}{c|}{\cellcolor[HTML]{EFEFEF}\better{5.53}} &
  2 &
  \multicolumn{1}{c|}{\cellcolor[HTML]{EFEFEF}\better{6.06}} &
  2.21 &
  \multicolumn{1}{c|}{\cellcolor[HTML]{EFEFEF}\better{5.84}} &
  4.04 &
  \better{6.88} \\
Benchmark 9 &
  16.39 &
  \multicolumn{1}{c|}{\cellcolor[HTML]{EFEFEF}16.39} &
  8.53 &
  \multicolumn{1}{c|}{\cellcolor[HTML]{EFEFEF}8.53} &
  7.35 &
  \multicolumn{1}{c|}{\cellcolor[HTML]{EFEFEF}\better{6.8}} &
  16.39 &
  16.39 &
  19.15 &
  \multicolumn{1}{c|}{\cellcolor[HTML]{EFEFEF}19.15} &
  7.69 &
  \multicolumn{1}{c|}{\cellcolor[HTML]{EFEFEF}7.69} &
  3.98 &
  \multicolumn{1}{c|}{\cellcolor[HTML]{EFEFEF}\better{6.41}} &
  16.39 &
  16.39 \\
Benchmark 10 &
  15.82 &
  \multicolumn{1}{c|}{\cellcolor[HTML]{EFEFEF}15.82} &
  11.23 &
  \multicolumn{1}{c|}{\cellcolor[HTML]{EFEFEF}11.23} &
  5 &
  \multicolumn{1}{c|}{\cellcolor[HTML]{EFEFEF}\better{9.83}} &
  15.82 &
  15.82 &
  10.57 &
  \multicolumn{1}{c|}{\cellcolor[HTML]{EFEFEF}10.57} &
  10.75 &
  \multicolumn{1}{c|}{\cellcolor[HTML]{EFEFEF}10.75} &
  4.94 &
  \multicolumn{1}{c|}{\cellcolor[HTML]{EFEFEF}\better{5.26}} &
  15.82 &
  15.82 \\
Benchmark 11 &
  24.27 &
  \multicolumn{1}{c|}{\cellcolor[HTML]{EFEFEF}24.27} &
  17.48 &
  \multicolumn{1}{c|}{\cellcolor[HTML]{EFEFEF}17.48} &
  9.69 &
  \multicolumn{1}{c|}{\cellcolor[HTML]{EFEFEF}9.69} &
  24.27 &
  24.27 &
  23.14 &
  \multicolumn{1}{c|}{\cellcolor[HTML]{EFEFEF}23.14} &
  14.04 &
  \multicolumn{1}{c|}{\cellcolor[HTML]{EFEFEF}14.04} &
  7.25 &
  \multicolumn{1}{c|}{\cellcolor[HTML]{EFEFEF}7.25} &
  24.27 &
  24.27 \\
Benchmark 12 &
  12.62 &
  \multicolumn{1}{c|}{\cellcolor[HTML]{EFEFEF}12.62} &
  15.01 &
  \multicolumn{1}{c|}{\cellcolor[HTML]{EFEFEF}15.01} &
  15.01 &
  \multicolumn{1}{c|}{\cellcolor[HTML]{EFEFEF}15.01} &
  13.55 &
  13.55 &
  17.3 &
  \multicolumn{1}{c|}{\cellcolor[HTML]{EFEFEF}17.3} &
  12.34 &
  \multicolumn{1}{c|}{\cellcolor[HTML]{EFEFEF}12.34} &
  14.04 &
  \multicolumn{1}{c|}{\cellcolor[HTML]{EFEFEF}14.04} &
  12.62 &
  12.62 \\ \hline
\end{tabular}
% \vspace{-0.2 in}
\end{table*}

\subsection{Mitigating BoS Vulnerabilities}
Algorithm~\ref{algo: HASTE-FIX} mitigates the BoS vulnerability in RSA modular exponentiation by synthesizing hardware-in-the-loop patched subroutines (Table~\ref{tab:benchmark_deatils}), where $\mathcal{S}$ denotes the branch on secret key bit $e$ and $\mathcal{N}_{Run}=100$. 
% To evaluate the effect of indirect branching time, an injected branch assigns 2 and 1 cycles to its two successors, $\mathcal{B}_{t1}$ and $\mathcal{B}_{t2}$, with the longest taken with probability $\mathcal{P}$ = 0.99.

\subsubsection{Baseline (No Cache/Branch Effect)} Fig.~\ref{B0S_5} illustrates the BoS severity for six benchmarks, $\mathcal{T}$ = 5 (marked by a red dashed line), comparing multiple devices before and after applying the \name~framework. Each subplot represents a specific benchmark, illustrating BoS severity levels, where higher values indicate a greater risk of information leakage. \name~reduces the severity value under the attacker's timing sensitivity for all pairs of benchmark devices except benchmark 2, which follows a similar trend, except for Devices 1 and 4, since the BoS severity of the benchmark within these devices is already below or equal to $\mathcal{T} = 5$.
% For Benchmark 1, devices initially exhibit high severity values for the vulnerable code (e.g., Device 2 at 16.1), but the framework of \name~significantly reduces these values closer to or below the threshold ($\mathcal T \leq 5$), such as Device 1 reducing from 12.8 to 3.6. Benchmark 2 follows a similar trend, except for Devices 1 and 4, as the the BoS severity of benchmark within those devices is already below or equal to the $\mathcal{T} =  5$. In Benchmark 3, all devices show a marked reduction in BoS severity, particularly Device 1, which drops from 10.5 to 1.5 which is below $\mathcal{T} \leq 5$. Benchmark 4 demonstrates consistent reductions in severity across all devices, bringing most values close to the acceptable threshold ($\leq 5$). In Benchmark 5, devices exhibit relatively lower severities compared to other benchmarks but still benefit from the \name~ as we can see in Device 1, the initial severity is very low (0.5) which means for Device 1, Benchmark 5 does not need to be fixed. Benchmark 6, all devices see substantial reductions with \name~, notably Device 5 reducing from 14.0 to 1.0, which is $\mathcal T \leq 5$. 
The differences between devices arise from the microarchitectural constant that remains even after flushing caches and removing the branching effects (e.g., pipeline depth/width and front-end bandwidth (fetch/decode), instruction-latency mix and micro code sequences, out-of-order and renaming resources, vector/ALU port availability, and clock frequency). Architectures like Intel’s i9-11900H (Cypress Cove), Cortex-A78AE, and Cortex-A76 have wider front ends (decode more instructions per cycle), larger out-of-order windows and register files, more execution ports, and shorter instruction latencies; Thus, wider cores issue/retire more micro-ops per cycle and hide latency better, completing each basic block in fewer cycles than narrower Cortex-A57, Carmel, or Cortex-A72 designs. Similar results are observed for $\mathcal{T}=10$ in Fig.~\ref{B0S_10}.

% , utilizing Algorithm~\ref{algo: HASTE-FIX}. 

\begin{takeaway}
\textbf{Takeaway}: \name~cuts BoS severity by up to \(\approx 93\%\) (e.g., \(12.8\to3.6\), \(10.5\to1.5\), \(14.0\to1.0\)), driving severities to \(\le \mathcal{T}=5\) with \(\mathcal{N}_{\text{Run}}=100\).
\end{takeaway}
% \vspace{-0.1in}
% The results, shown in Figure~\ref{B0S_10}, reveal that \name~is precisely mitigating the timing side channel considering the underlying hardware architecture.

\subsubsection{Considering the Effect of Branch Prediction} 
% Branching effects refer to the solution, as a plain CFG cannot dissolve runtime information (indirect branches). 
The left half of the Table~\ref{branching_and_Cache_effect} reports BoS severity for the source code and after applying \name~when the attacker timing threshold $\mathcal{T}$ = 10.  Almost all the benchmarks show a decreasing trend in BoS severity values. Several entries did not change, for example, Intel Core i9 Benchmarks 5 (3.37), 6 (10.00), 9 (3.05), and 11 (2.06) because the original paths were already symmetric or below the threshold $\mathcal{T}$=10.
% so no padding was applied. 
\vspace{-0.05in}
\begin{takeaway}
    \textbf{Takeaway}: With attacker threshold \(\mathcal{T}=10\), \name~cuts BoS severity on average \(\approx61\%\) across multiple devices; Benchmarks like B5/B6/B9/B11 remain same since the severity is already \(\leq\mathcal{T}\).
\end{takeaway}
\vspace{-0.05in}
BoS value differ across hardware because the measured per-block cycle numbers differ across hardware (e.g. different pipelines, execution latencies, clocking). Even with the same $\mathcal{B}_{t1}$,  $\mathcal{B}_{t2}$ and $\mathcal{P}$ for all pairs of benchmark devices.

\subsubsection{Considering Both Cache State \& Branch Prediction Effect} The right side of the Table~\ref{branching_and_Cache_effect} extends the control-flow model to include the effect of different cache states. BoS declines for most device–benchmark pairs; Benchmarks 5, 9, 10, 11, and 12 are unchanged because they already meet the LoS resilience threshold.
% Because cache hits and misses, as well as high-latency operations (e.g., division), add timing variability, the original (Orig) severities are typically larger.\name~compresses the gap by making the two paths timing almost similar by using non-operational padding. 
% On Core i9, severity falls below the threshold ($\mathcal T \leq10$), for example, benchmark 2 drops from 19.4 to 2.3, and benchmark 3 from 14.67 to 3.50. Jetson Nano shows a similar trend (Benchmark 1 decreases from 19.0 to 9.56), as do the Raspberry Pi (Benchmark 3 from 18.98 to 8.05) and Jetson AGX (Benchmark 3 from 13.9 to 4.12). 
The differences across devices arise from how each device’s memory system is built. Cache size/associativity (whether data fits or spills), cache hit/miss latency, prefetcher behavior, TLB coverage, and DRAM bandwidth all vary by platform. Because of this, the same uneven access pattern may cause many cache/TLB misses on one CPU but few on another, so the timing penalty can be large on one device and small on another.
% \vspace{-0.1in}
\begin{takeaway}
     \textbf{Takeaway}: Accounting for cache states, \name's padding cuts BoS severity by \(\sim 50\text{--}88\%\) across platforms  when, \(\mathcal{T}\leq 10\).

\end{takeaway}

% Overall, the \name~ framework effectively mitigates BoS Severity across all devices and benchmarks, demonstrating its ability to reduce timing vulnerabilities and bring severity levels closer to or below the designated threshold. This highlights its utility in enhancing security in systems prone to branch-related information leakage.
% \vspace{-0.1in}

\subsection{\name~for LoS Mitigation}
We use Algorithm~\ref{algo: HASTE-FIX-Los} to mitigate LoS (Loop on Secret) vulnerabilities by generating new software subroutines via hardware-in-the-loop testing. The \name~ framework was evaluated at a timing sensitivity ($\mathcal{T}$) = 50 with, $\mathcal{N}_{Run}$ = 100, $\mathcal{B}_{t1}$ = 2, $\mathcal{B}_{t2}$ = 1 and $\mathcal{P}$ = 0.99. The goal is to maximize LoS resilience, which measures the minimum loop iterations needed to detect timing discrepancies, with higher values indicating greater security. 

\subsubsection{Baseline (No Cache/Branch Effect)}Fig.~\ref{LOS_10} shows LoS resilience across benchmarks on five devices. Each subplot corresponds to one device, with benchmarks on the x-axis and LoS resilience on the y-axis; the red dashed line marks the minimum threshold, $\mathcal{L}_{\min}=10$.

\begin{takeaway}
    \textbf{Takeaway}: \name~boosts LoS resilience by pushing it above the \(\mathcal{L}_{min}=10\) across most benchmarks in 4/5 devices (80\%).
\end{takeaway}

% For Device 1, we see a significant spike in LoS resilience for one benchmark, reaching a value of 100, while resilience remains relatively low for other benchmarks. \name~does not drastically alter resilience to maintain consistency with improvement around the threshold, for example, in Device 1 for Benchmark 1, LoS resilience is 20, which is already above the minimum LoS resilience $(5)$. 
Device 1, Device 2, Device 3, Device 4, and Device 5 demonstrate a notable improvement in resilience value with \name~for certain benchmarks, effectively surpassing the threshold $(\geq 10)$ in most cases, whereas resilience under normal conditions varies more significantly. \name~does not apply any fixes if the device-benchmark pair resilience is already satisfactory (per threat model). For example, in Device 1 for Benchmark 1, LoS resilience is 20, which is already above the minimum LoS resilience $10$. A similar sets of results is presented in Fig.\ref{LOS_5} when LoS resilience is 5.

% For Device 2, vulnerable benchmarks' LoS Resilience values fluctuate below and around the threshold ($\mathcal{L}_{min} = 5$) across benchmarks, whereas the use of \name~results in consistent improvement, pushing resilience above the threshold ($\mathcal{L}_{min} = 5$). Device 3, Device 4, and Device 5 demonstrate a notable improvement in resilience value with \name~for certain benchmarks, effectively surpassing the threshold $(\geq 5)$ in most cases, whereas resilience under normal conditions varies more significantly.
% We consider a similar experiment using $\mathcal{L}_{min}$ = 10 using Algorithm~\ref{algo: HASTE-FIX-Los}. The results are illustrated in Figure~\ref{LOS_10}.

\begin{figure}[!htb]
\centering

\includegraphics[width=\columnwidth]{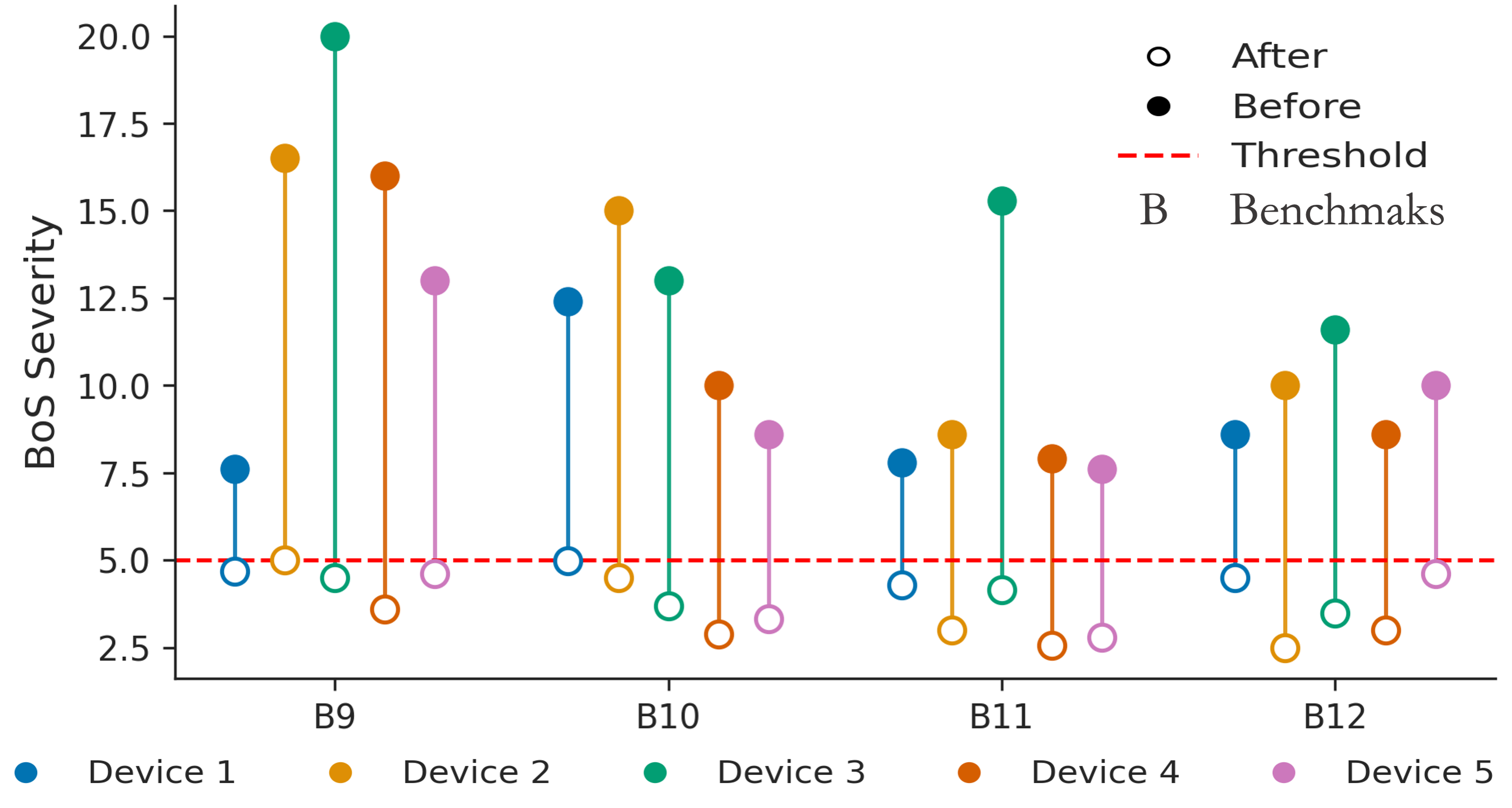}
\captionsetup{justification=centering}
\caption{BoS severity (before and after applying \name) for Traces ($\mathcal{N_T}$) = 10 and Timing sensitivity $\mathcal{T}$ = 5. \label{fig:traces}}

\vspace{-0.2in}
\end{figure}
\begin{figure}[!htb]
\centering

\includegraphics[width=\columnwidth]{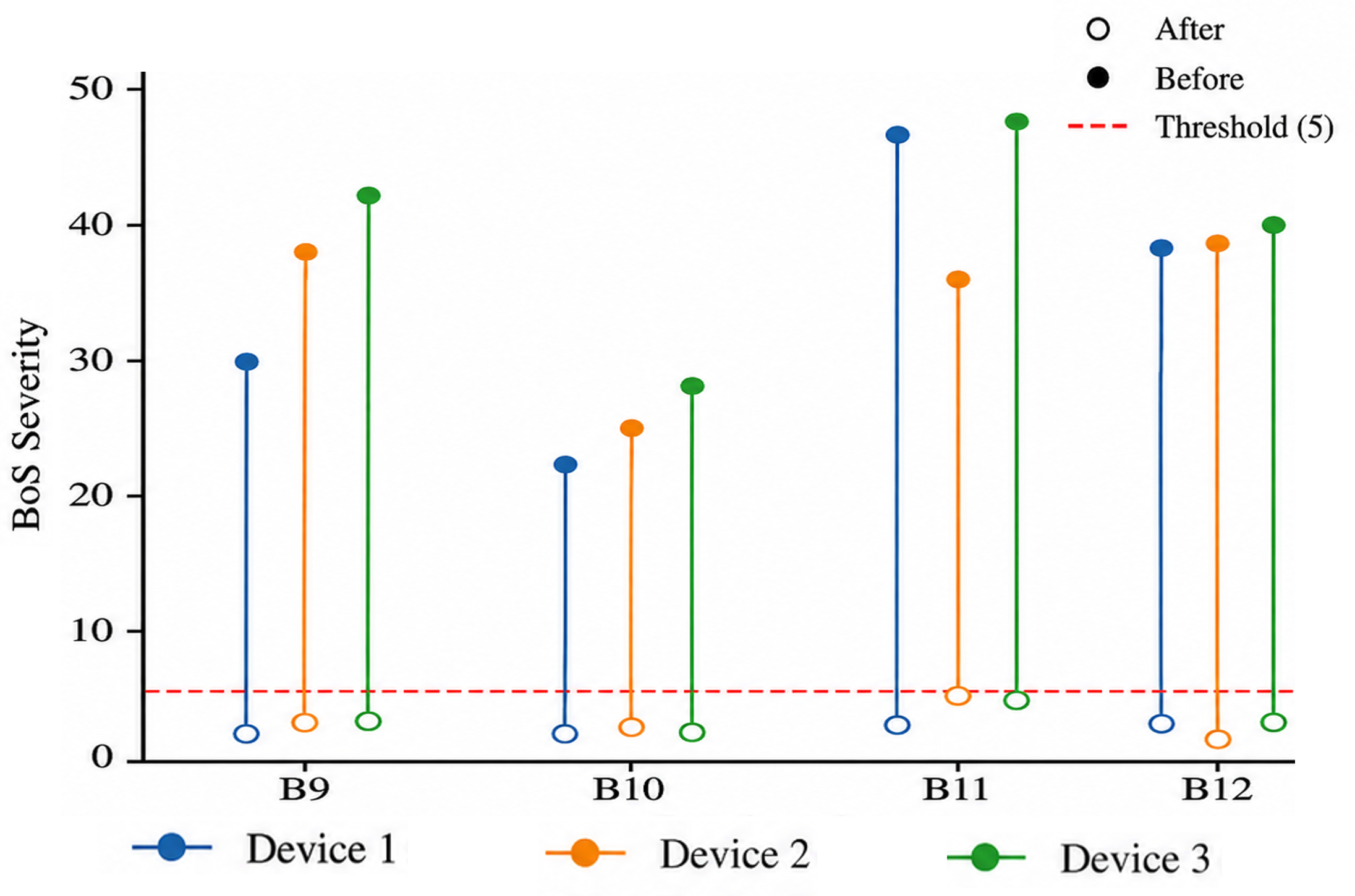}
\captionsetup{justification=centering}
\caption{BoS severity (before and after applying \name) for Traces ($\mathcal{N_T}$) = 100 and Timing sensitivity $\mathcal{T}$ = 5.} \label{fig:traces_100}

% \vspace{-0.2in}
\end{figure}

\subsubsection{Considering the Effect of Branch Prediction} The left half of the Table~\ref{branching_and_Cache_effect_LOS} reports LoS severity for the original source code and after applying \name~when $\mathcal{L}_{min}=5$. Our framework \name~is able to approximate the effect of branch prediction given the specified user-defined parameters, $\mathcal{B}_{t1}$, $\mathcal{B}_{t2}$, and $\mathcal{P}$. We consistently observe improved LoS resilience across diverse benchmark--device combinations.

% n Intel Core i9, several benchmarks move from below 5 to comfortably above it. For example, Benchmark 1 increases from 3.63 to 6.71, Benchmark 3 from 4.05 to 14.28, Benchmark 4 from 4.96 to 8.15, and Benchmark 7 from 3.65 to 8.53. On Jetson Nano, most sub-5 entries are raised above the threshold: Benchmark 1 rises from 2.71 to 5.34, Benchmark 2 from 4.16 to 6.47, Benchmark 3 from 3.75 to 5.91, Benchmark 4 from 3.35 to 6.38, Benchmark 6 from 2.97 to 5.26, Benchmark 7 from 2.60 to 5.20, and Benchmark 8 from 2.06 to 6.41. Raspberry Pi already has many benchmarks at or near the target, with notable increases such as Benchmark 2 from 5.10 to 9.75 and Benchmark 6 from 4.5 to 10.00; On Jetson AGX, the pattern is similar to Core i9 with moderate lifts (e.g., Benchmark 1 from 3.63 to 6.71 and Benchmark 4 from 4.96 to 6.14). Entries that remain unchanged, such as Core i9 benchmarks 5, 9-11 and Nano benchmarks 9-12, were already at or well above $\mathcal{L}_{min}$, so further lifting was unnecessary.
\begin{takeaway}
    \textbf{Takeaway}: Applying \name~ raises the minimum LoS resilience in all devices with gains of \(64\%\to 253\% \) (core i9), \(56\%\to211\%\) 
    (Jetson Nano), \(91\%\to122\%\) (Raspberry Pi) and \(24\%\to85\%\) (Jetson AGX), while the entries already above the threshold remain unchanged.
\end{takeaway}

\subsubsection{Considering Both Cache State \& Branch Prediction Effect} The right half of the Table~\ref{branching_and_Cache_effect_LOS} extends the control-flow model to include the effect of different cache states. 
% On Intel Core i9, Benchmark 2 jumps from 2.57 to 21.73 and Benchmark 3 from 3.40 to 14.28, while several others see smaller but still meaningful gains (e.g., Benchmark 4 from 4.22 to 5.17). On Jetson Nano, a broad set of benchmarks crosses the  $\mathcal{L}_{min}$ line: Benchmark 1 rises from 2.63 to 5.23; Benchmark 2 from 4.04 to 5.75; Benchmark 3 from 3.57 to 5.50; Benchmark 4 from 3.30 to 5.47; Benchmark 6 from 2.89 to 5.05; Benchmark 7 from 2.60 to 5.11; and Benchmark 8 from 2.00 to 6.06. Raspberry Pi shows similar behavior, with improvements such as Benchmark 1 from 4.16 to 7.09, Benchmark 2 from 3.10 to 5.50, Benchmark 3 from 2.63 to 6.21, Benchmark 4 from 2.77 to 7.35, Benchmark 7 from 1.85 to 5.48, Benchmark 8 from 2.21 to 5.84, and Benchmark 9 from 3.98 to 6.41. On Jetson AGX, the most visible gains are Benchmark 1 from 3.59 to 6.06, Benchmark 3 from 3.59 to 12.13, Benchmark 4 from 4.96 to 6.13, Benchmark 7 from 3.58 to 8.53, and Benchmark 8 from 4.04 to 6.88; several higher resilience entries (e.g., Benchmarks 5, 9–11, 12) stay unchanged because they already met the target. 
Results across devices reflect the differences in cache capacity, associativity, hit/miss latencies, prefetcher, and DRAM bandwidth.
% \vspace{-0.3in}
\begin{takeaway}
    \textbf{Takeaway}: Accounting for cache states, the LoS metric roughly doubles on average (median \(\times 1.9\)) and passes \(\mathcal{L}_{\min}\), which means that the LoS severity is mitigated for the benchmarks across multiple devices.
\end{takeaway}

\vspace{-0.5cm} 
\subsection {\name~for Mitigating Multiple Timing Trace Attacks}

A powerful timing attacker can improve their success rate by collecting repeated measurements from the same target. We therefore set the timing sensitivity to $\mathcal{T}=5$ and evaluate $\mathcal{N_T}=10,100$ traces. We choose four benchmarks (Benchmarks 9--12), which are vulnerable before fixing at the same sensitivity on five heterogeneous devices (three devices for $\mathcal{N_T}=100$). As shown in Figs.~\ref{fig:traces} and~\ref{fig:traces_100}, all benchmarks initially violate the threshold, with leakage ranging from \texttt{$\sim$}20 cycles in Feistel on Device~3 to \texttt{$\sim$}9 cycles in Hill on Device~1. After applying \name~via Algorithm~\ref{algo: HASTE-FIX}, all benchmark--device pairs fall within 2.5--5 cycles.
% which confirms that \name~ can also secure the system, if the attacker is able to collect multiple traces. 

\vspace{-0.1in}

\section{Discussion}
\subsection{Limitations of \name~}
\name~currently mitigates timing leakage by inserting semantically neutral dummy operations through \texttt{AddNoise()}. This is lightweight and effective for the timing-focused BoS/LoS threat model considered in this work. However, the current implementation uses a fixed padding pattern, which may be easier to distinguish under other observation channels, such as power, electromagnetic, or fault-based attacks. Therefore, deployments targeting these stronger adversaries may require diversified padding sequences. The current implementation supports structured control flow, including straight-line code, if/else branches, for/while loops, arithmetic operations, and common array accesses. It does not claim full support for unrestricted recursion, indirect jumps, \texttt{goto} statements, or function-pointer-based control flow. When such constructs are detected, \name~reports the region as unsupported rather than applying an unsafe automatic repair.

\subsection{Future Research Directions}
As future work, we will diversify \texttt{AddNoise()} with semantically neutral dummy-operation sequences, making padding less predictable under non-timing channels such as power, electromagnetic, or fault-based attacks while preserving the existing BoS/LoS detection and repair logic. We will also extend \name~to support more complex control flow, including recursion, indirect jumps, and function-pointer dispatch, and evaluate it under stronger adversarial settings such as attacker-controlled cache/branch-predictor states and worst-case leakage-maximizing input pairs. Finally, we will improve scalability through parallel module/procedure-level CFG construction, compact interprocedural summaries inspired by ThinLTO~\cite{johnson2017thinlto}, and CPU/GPU-based sparse-graph traversal frameworks such as Hornet~\cite{busato2018hornet}.

% \vspace{-0.1in}
\section{Conclusion}
Timing side-channel attacks can expose sensitive data from embedded systems that are routinely used in a wide range of applications (e.g. healthcare, additive manufacturing). State-of-the-art software runtime side-channel attack mitigation frameworks do not consider the underlying hardware information leading to over/under-fixing of the vulnerabilities. In this work, we introduce an automated framework (\name) that can mitigate software timing vulnerabilities with a complete transparent understanding of the underlying electronic hardware and the associated threat model. We have integrated \name~ into the commercial embedded systems development flow and demonstrate its effectiveness in detecting/mitigating timing-based leakage for 22 software benchmarks across five unique embedded devices. \name~ was able to optimally mitigate all discovered vulnerabilities with lower overhead compared to constant-time solutions. 
%Future works will extend \name~ for other side channel threats (e.g. power, EM).    
% \vspace{-0.5cm}
\section{Acknowledgment}
The authors gratefully acknowledge funding and technical support from the U.S. Army Engineer Research and Development Center ITL via Other Transaction Agreement \#W15QKN-17-9-5555 Sub-Agreement \#C5-23-1003.

%%%%%%%%%%%%%%%%%%%%%%%%%%%%%%%%%%%%%%%%%%%%%%%%%%%%%%%%%%%%%%%%%%%%%%%%%%%%%%%%

%%%%%%%%%%%%%%%%%%%%%%%%%%%%%%%%%%%%%%%%%%%%%%%%%%%%%%%%%%%%%%%%%%%%%%%%%%%%%%%%

%%%%%%%%%%%%%%%%%%%%%%%%%%%%%%%%%%%%%%%%%%%%%%%%%%%%%%%%%%%%%%%%%%%%%%%%%%%%%%%%

%%%%%%%%%%%%%%%%%%%%%%%%%%%%%%%%%%%%%%%%%%%%%%%%%%%%%%%%%%%%%%%%%%%%%%%%%%%%%%%%

% \clearpage

\bibliographystyle{IEEEtran}

\bibliography{IEEEabrv,SOCC}

\end{document}